\documentclass[aps,prl,twocolumn,showpacs,showkeys,groupedaddress]{revtex4-1}

\usepackage{amsmath}
\usepackage{graphicx}
\usepackage{color}
\usepackage{amssymb} 
\usepackage{ulem}
\usepackage{hyperref}

\def\TeV{\ifmmode {\mathrm{Te\kern -0.1em V}}\else
                   \textrm{Te\kern -0.1em V}\fi}%
\def\GeV{\ifmmode {\mathrm{Ge\kern -0.1em V}}\else
                   \textrm{Ge\kern -0.1em V}\fi}%
\def\MeV{\ifmmode {\mathrm{Me\kern -0.1em V}}\else
                   \textrm{Me\kern -0.1em V}\fi}%
\def\keV{\ifmmode {\mathrm{ke\kern -0.1em V}}\else
                   \textrm{ke\kern -0.1em V}\fi}%
\def\eV{\ifmmode  {\mathrm{e\kern -0.1em V}}\else
                   \textrm{e\kern -0.1em V}\fi}%

\let\gev=\GeV

\def\MSbar{\mbox{$\overline{\mathrm{MS}}$\,}}% MS bar scheme

\begin{document}

% Use the \preprint command to place your local institutional report
% number in the upper righthand corner of the title page in preprint mode.
% Multiple \preprint commands are allowed.
% Use the 'preprintnumbers' class option to override journal defaults
% to display numbers if necessary
%\preprint{}

%Title of paper
\title{Snowmass White Paper: prospects for measurements of the bottom quark mass}

% repeat the \author .. \affiliation  etc. as needed
% \email, \thanks, \homepage, \altaffiliation all apply to the current
% author. Explanatory text should go in the []'s, actual e-mail
% address or url should go in the {}'s for \email and \homepage.
% Please use the appropriate macro foreach each type of information

% \affiliation command applies to all authors since the last
% \affiliation command. The \affiliation command should follow the
% other information
% \affiliation can be followed by \email, \homepage, \thanks as well.

\author{Javier Aparisi}
%\email[]{Your e-mail address}
%\altaffiliation{}
\affiliation{Instituto de F\'isica Corpuscular, CSIC-University of Valencia, Valencia, Spain}

\author{Juan Fuster}
%\email[]{Your e-mail address}
%\altaffiliation{}
\affiliation{Instituto de F\'isica Corpuscular, CSIC-University of Valencia, Valencia, Spain}

\author{Andr\'e Hoang}
\affiliation{University of Vienna, Austria}

\author{Adri\'an Irles}
%\email[]{Your e-mail address}
\affiliation{Instituto de F\'isica Corpuscular, CSIC-University of Valencia, Valencia, Spain}
%\affiliation{LAL, Orsay, France}

\author{Christopher Lepenik}
\affiliation{University of Vienna, Austria}

\author{Vicent Mateu}
\affiliation{Departamento de F\'isica Fundamental e IUFFyM, Universidad de Salamanca, Salamanca, Spain}

\author{Germ\'an Rodrigo}
\altaffiliation{European Research Council Executive Agency, European Commission, BE-1049 Brussels, Belgium. Disclaimer: the views expressed in this article are strictly those of the author and may not in any circumstance be regarded as stating an official position of the European Commission.}
\affiliation{Instituto de F\'isica Corpuscular, CSIC-University of Valencia, Valencia, Spain}

\author{Michael Spira}
\affiliation{PSI Villigen, Switzerland}

\author{Seidai Tairafune}
%\email[]{Your e-mail address}
%\altaffiliation{}
\affiliation{Tohoku University, Sendai, Japan}

\author{Junping Tian}
\affiliation{University of Tokyo, Tokyo, Japan}

\author{Marcel Vos}
\email[]{marcel.vos@ific.uv.es}
\affiliation{Instituto de F\'isica Corpuscular, CSIC-University of Valencia, Valencia, Spain}

\author{Hitoshi Yamamoto}
%\email[]{Your e-mail address}
\altaffiliation{On leave from Tohoku University, Sendai, Japan}
\affiliation{Instituto de F\'isica Corpuscular, CSIC-University of Valencia, Valencia, Spain}

\author{Ryo Yonamine}
%\email[]{Your e-mail address}
%\altaffiliation{}
\affiliation{Tohoku University, Sendai, Japan}

%Collaboration name if desired (requires use of superscriptaddress
%option in \documentclass). \noaffiliation is required (may also be
%used with the \author command).
%\collaboration can be followed by \email, \homepage, \thanks as well.
%\collaboration{}
%\noaffiliation

\date{\today}

\begin{abstract}
  In this white paper for the Snowmass '21 community planning exercise we provide quantitative prospects for bottom quark mass measurements in high-energy collisions at future colliders that can provide a precise test of the scale evolution, or ``running'' of quark masses predicted by QCD.
\end{abstract}

% insert suggested PACS numbers in braces on next line
% insert suggested keywords - APS authors don't need to do this

%\maketitle must follow title, authors, abstract, \pacs, and \keywords
\maketitle

% body of paper here - Use proper section commands
% References should be done using the \cite, \ref, and \label commands
\section{Introduction}
% Put \label in argument of \section for cross-referencing
%\section{\label{}}

%After renormalization, the QCD Lagrangian Eq. (66.1) gives finite values
%for physical quantities, such as scattering amplitudes. Renormalization is a procedure
%that invokes a subtraction scheme to render the amplitudes finite, and requires the
%introduction of a dimensionful scale parameter μ. The mass parameters in the QCD
%Lagrangian Eq. (66.1) depend on the renormalization scheme used to define the theory,
%and also on the scale parameter μ. The most commonly used renormalization scheme for
%QCD perturbation theory is the MS scheme.

The masses of quarks are free parameters in the Standard Model of particle physics, whose values must be determined experimentally. Precise measurements are performed through the comparison of measurements of physical observables sensitive to the mass to SM predictions in Quantum Chromodynamics (QCD) beyond leading order accuracy. Like the strong coupling constant, quark masses depend on the renormalization scheme and are renormalization-scale-dependent parameters or ``running constants''. In this study, we adopt the most popular renormalization scheme, the modified minimal subtraction scheme or $\overline{\rm MS}$ scheme, where the strong coupling $\alpha_s(\mu)$ and the quark masses $m_q(\mu)$ depend on the dimensionful renormalization scale $\mu$ that is identified with the energy scale of the scattering process~\footnote{Throughout this White Paper we define the $\overline{\rm MS}$ bottom quark mass in the $n_f=5$ flavor scheme.}. The scale evolution is predicted by QCD. Given a measurement at one scale, the value at any other scale is predicted by the renormalization group equation (RGE). RGE calculations have reached 5-loop (${\cal O}(\alpha_s^5)$) accuracy~\cite{Vermaseren:1997fq,Chetyrkin:1997dh,Baikov:2014qja} and software packages such as RunDec~\cite{Herren:2017osy} and REvolver~\cite{Hoang:2021fhn} provide access to state-of-the-art renormalization evolution and scheme conversions. The predicted evolution can be tested experimentally by performing measurements at different scales.
The evolution of the strong coupling has been tested over a broad range of energies~\cite{Zyla:2020zbs} and experiments have studied the ``running'' of \MSbar{} quark masses for the the charm quark at HERA~\cite{Gizhko:2017fiu} and the top quark at the LHC~\cite{Sirunyan:2019jyn}. We discuss the case of the bottom quark mass in this White Paper.

The most precise extractions of the bottom quark mass~\cite{Narison:2019tym,Peset:2018ria,Kiyo:2015ufa,Penin:2014zaa,Alberti:2014yda,Beneke:2014pta,Dehnadi:2015fra,Lucha:2013gta,Bodenstein:2011fv,Laschka:2011zr,Chetyrkin:2009fv} rely on the measurement of the mass of bottomonium bound states and the $e^+e^- \rightarrow $ hadrons cross section as experimental input, in combination with QCD sum rules and perturbative QCD calculations. Several lattice QCD groups have also published results, the most recent of which reaches a precision of approximately 0.3\%~\cite{Bazavov:2018omf,Colquhoun:2014ica,Bernardoni:2013xba,Lee:2013mla,Dimopoulos:2011gx} (see also the FLAG report~\cite{Aoki:2019cca}). The world average provided by the Particle Data Group (PDG)~\cite{Zyla:2020zbs} also includes inputs from HERA~\cite{H1:2018flt} and the BaBar and Belle experiments at the B-factories~\cite{Schwanda:2008kw,Aubert:2009qda}. This world average of low-scale bottom-quark mass measurements has a relative precision better than 1\%:
\begin{equation}
  m_b(m_b) = 4.18^{+0.03}_{-0.02}~\gev{},
\label{eq:mbmb}
\end{equation}
where the reference value of the bottom mass is quoted in the \MSbar\ scheme, at a scale given by the mass itself. 

Bottom quark mass measurements at a much higher scale became possible at LEP and SLC, where jet rates and event shapes are sensitive to subleading mass effects. A practical method to extract the bottom-quark mass from $Z$-pole data was proposed in Ref.~\cite{Bilenky:1994ad}. Three independent groups completed the necessary next-to-leading order (NLO) theoretical calculation of the three-jet rate for massive quarks~\cite{Rodrigo:1997gy,Bilenky:1998nk,Rodrigo:1999qg,Bernreuther:1997jn,Brandenburg:1997pu,Nason:1997tz,Nason:1997nw} (an NNLO calculation for the three-jet rate in $e^+e^-$ collisions, without bottom-quark-mass effects, is available in Ref.~\cite{GehrmannDeRidder:2008ug}). The first measurement of this type was performed by the DELPHI collaboration~\cite{Abreu:1997ey} using the LEP $Z$-pole data. Similar measurements were also performed with SLD~\cite{Brandenburg:1999nb,Abe:1998kr} data, and by ALEPH~\cite{Barate:2000ab}, OPAL~\cite{Abbiendi:2001tw} and DELPHI~\cite{Abdallah:2005cv,Abdallah:2008ac}. The combination of the most precise determinations from three-jet rates of each experiment yields the following average for $m_b(m_Z)$:
\begin{equation}
    m_b(m_Z) = 2.82 \pm 0.28~\GeV{}.
    \label{eq:mbmz}
\end{equation}
This value is in good agreement with the average of Ref.~\cite{Kluth:2006bw} that is based on a slightly different sub-set of measurements.
 
 Recently, a new measurement was published of the bottom quark mass at the scale of the Higgs boson mass~\cite{Aparisi:2021tym}. The value of $m_b(m_H)$ is inferred from measurements by the ATLAS~\cite{ATLAS-CONF-2020-027} and CMS~\cite{Sirunyan:2018koj} experiments of the bottom quark decay width to bottom quarks $\Gamma(H\rightarrow b\bar{b}$). The width is normalized to $\Gamma(H\rightarrow ZZ)$, the decay width for the $ZZ$ decay mode. The average of the $m_b(m_H)$ results obtained from both measurements yields:
\begin{equation}
    m_b(m_H) = 2.60 ^{+0.36}_{-0.30}~\GeV{}.
    \label{eq:mbmh}
\end{equation}
This result reinforces the experimental evidence for the ``running'' of the \MSbar{} bottom quark mass, definitively excluding the no-running scenario with a statistical significance greater than 7 standard deviations. 

 \begin{figure}[h!]
\includegraphics[width=0.99\columnwidth]{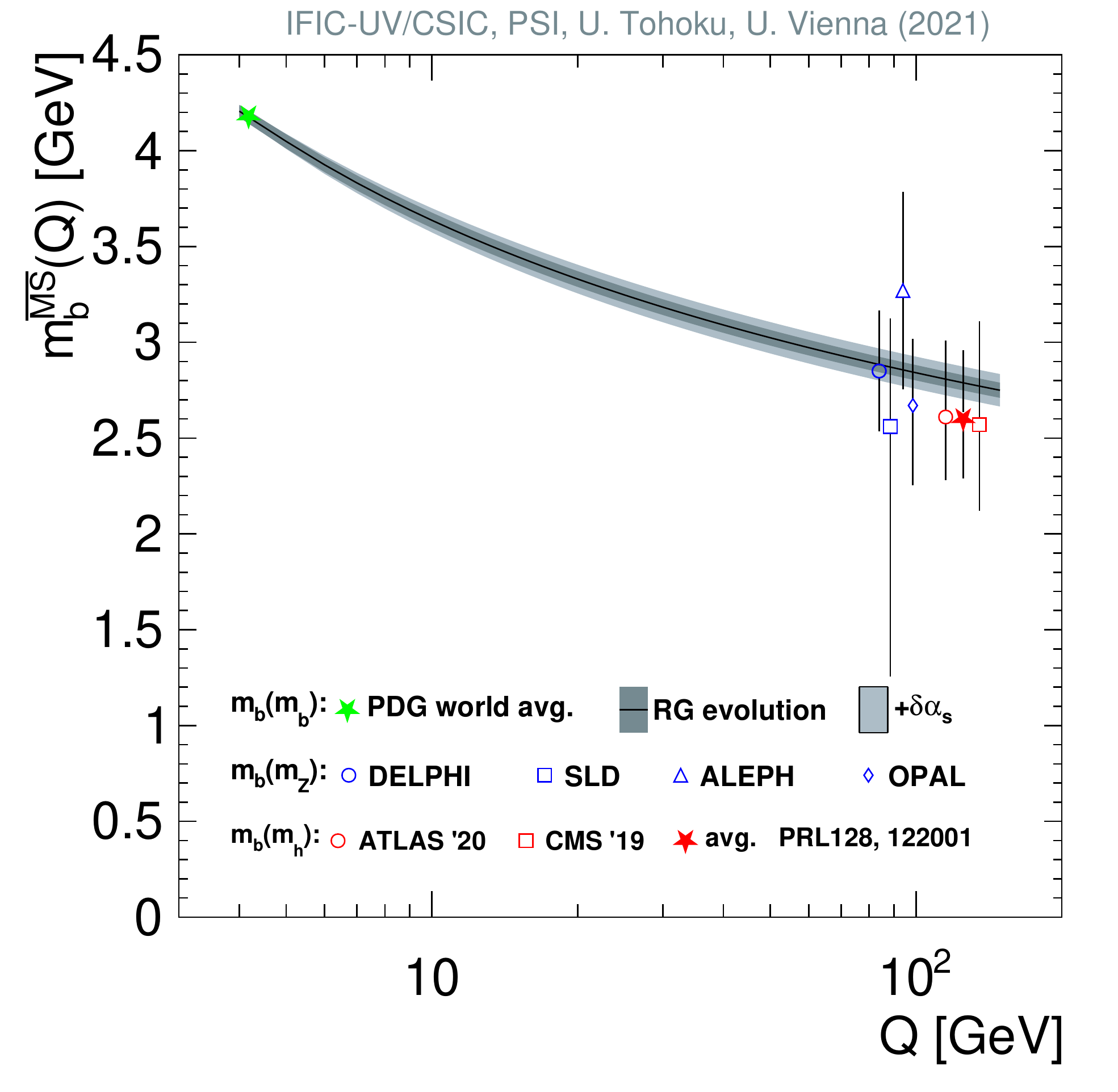}%
\caption{\label{fig:running_mass} The scale evolution of the bottom quark \MSbar{} mass. The measurements include the PDG world average for $m_{b}(m_{b})$ from low-scale measurements, the measurements of $m_b(m_Z)$ from jet rates at the \mbox{$Z$-pole} at LEP and SLC and the measurement of $m_b(m_H)$ from Higgs boson branching fractions. The prediction of the evolution of the mass is calculated at five-loop precision with REvolver~\cite{Hoang:2021fhn}. The inner dark grey error band includes the effect of missing higher orders and the parametric uncertainties from $m_b(m_b)$ and $\alpha_s$ from the PDG averages. The outer band with a lighter shading includes additionally the effect of a $\pm$ 0.004 variation of $\alpha_s(m_Z)$. Figure from Ref.~\cite{Aparisi:2021tym}.}
\end{figure}

The three sets of measurements, of $m_b(m_b)$, $m_b(m_Z)$ and $m_b(m_H)$, are shown in Fig.~\ref{fig:running_mass} and compared to the evolution of the PDG world average from $m_b(m_b)$ to a higher scale using the RGE calculation included in the REvolver code~\cite{Hoang:2021fhn} at five-loop precision.

We argue that in the next decade the study of the ``running'' of the bottom quark mass will turn into a precise test of QCD. Precise measurements at several energy scales can be used to rule out or confirm the presence of massive new coloured states that may contribute to the quark mass evolution. New collider facilities can further enhance this potential in several ways. In this White Paper we provide a quantitative assessment of the potential of the High-Luminosity phase of the LHC (HL-LHC) and a future ``Higgs factory" electron-positron collider to test the scale evolution of the bottom quark mass predicted by QCD.

\section{The bottom-quark mass from low-energy measurements}
Both from the theoretical and experimental points of view, the ideal observable to determine $m_b$ is the bottomonium spectrum. The masses of $b\bar b$ bound states are very sensitive to the bottom quark mass and have nearly vanishing experimental uncertainties. On the other hand, we can compute the masses of the low-lying narrow bottominum resonances only through perturbative expansions supplemented with non-perturbative corrections. The typical dynamical scale of these narrow bottominum masses are of the order of the inverse Bohr radius $\sim C_F\alpha_s m_b\gg \Lambda_{\rm QCD}$,
such that perturbative expansions can be used reliably and non-perturbative corrections remain small. Quarkonium masses have been computed to
%This translates into somewhat slowly converging perturbative series and enhanced non-perturbative effects (which may not even follow a local operator product expansion). {\color{red}[This all sounds totally weird and confusing. It is unclear which message on the reliability of the results this will send to the reader.]} These drawbacks are partially compensated by the fact that perturbative expansions are known to
N$^3$LO for arbitrary quantum numbers and ultrasoft resummation is known to N$^2$LL.
The most up to date analyses have been carried out in Refs.~\cite{Peset:2018ria} and \cite{Mateu:2017hlz} (earlier analyses use the $\overline{\rm MS}$ mass and are not discussed). They share some features, like employing low-scale short-distance masses (RS and MSR, respectively), using a 3-flavor scheme plus finite charm mass corrections and varying two renormalization scales, whose variation range is inferred from the perturbative logarithms. There are small differences as well:~while \cite{Peset:2018ria} determines the mass from a single bound state, \cite{Mateu:2017hlz} performs global fits with correlated scale variation using a $\chi^2$ function. In this latter analysis non-perturbative effects are estimated by comparing the $m_b$ results for different sets of quarkonia states.
%On the other hand \cite{Peset:2018ria} includes N$^3$LL ultrasoft resummation, which is only known for P-wave states.
The analyses in Ref.~\cite{Peset:2018ria} and~\cite{Mateu:2017hlz} both use PDG~\cite{Zyla:2020zbs} data and find compatible results. Ref.~\cite{Peset:2018ria} finds:
\begin{equation}
  m_b(m_b)= 4.186\pm0.037\,\rm GeV,
\end{equation}
while Ref.~\cite{Mateu:2017hlz} finds:
\begin{equation}
  m_b(m_b)= 4.216\pm0.039\,\rm GeV.
\end{equation}
Since experimental data is already extremely accurate, more precision in the future
could only come from the theory side. If a new perturbative order becomes available
the theoretical uncertainty (that clearly dominates) could go down to $0.026\,$GeV.
This estimate is obtained with a quadratic extrapolation (assuming that the error
from N$^3$LO to N$^4$LO goes down by the same factor as it does for going from
N$^2$LO to N$^3$LO yields a slightly smaller error). This should be taken
as a rough estimate only. On the other hand, if N$^3$LL resummation becomes
available for all bound states (so far it is only known for P-wave states), an additional error reduction should be expected as well.

Weighted averages of the bottom-tagged hadronic \mbox{R-ratio} in $e^+e^-$ annihilation are alternative physical observables to measure $m_b$. They are very sensitive to the bottom quark mass, and precise theoretical calculations can be achieved. This method goes under the name of QCD sum rules, and it can take different forms depending on the specific weight function and on whether the integration is cut off at some finite energy or not. For relativistic sum rules it was shown in Refs.~\cite{Dehnadi:2011gc,Dehnadi:2015fra} that in order to properly estimate perturbative uncertainties it is mandatory to vary the renormalization scales of the strong coupling $\alpha_s$ and bottom quark ${\overline {\rm MS}}$ mass independently, in order to avoid a biased dependence on the expansion prescription. On the experimental side one has to account for contributions from narrow resonances (whose masses and electronic widths can be taken from the PDG~\cite{Zyla:2020zbs}) and continuum data (so far only BaBar~\cite{BaBar:2008cmq} has released data), which only exists for energies below $11.21$\,GeV. This limits the experimental precision of this method if sum rules without an energy cut are considered (relativistic sum rules). These are very clean, theoretically, since their intrinsic dynamical scale is of order $m_b$. Alternatively, one can consider sum rules with an energy cut at the last experimental data point (finite-energy sum rules), or use moments with specific weight functions that strongly suppress the contributions from high-energies (non-relativistic sum rules). These approaches, in turn, have additional theoretical drawbacks, related to either a higher sensitivity to non-perturbative effects or to somewhat lower renormalization scales (as is also the case for the bottomonium mass method mentioned above).

We show results from two representative analyses for relativistic sum rules.
Ref.~\cite{Chetyrkin:2009fv} uses correlated scale variation for the bottom quark mass and $\alpha_s$ and finds:  
\begin{equation}
  m_b(m_b) = 4.163\pm0.016\,\rm GeV.
\end{equation}
Ref.~\cite{Dehnadi:2015fra} with the independent variation mentioned above, finding:
 \begin{equation} 
    m_b(m_b)= 4.176\pm0.023\,\rm GeV.
\end{equation}

where the experimental systematic uncertainty by itself amounts to $19$\,MeV and the theoretical uncertainty to $9$\,MeV for the more conservative analysis of Ref.~\cite{Dehnadi:2015fra}.
This uncertainty can be reduced substantially if additional data in the continuum region around and above $11.2$\,GeV becomes available. This would potentially open up the possibility for using
the $n=1$ moment, which is theoretically the cleanest and most precise, but has the strongest sensitivity to high-energy data. Assuming that this new data decreases the uncertainty on the continuum by a factor of $2$,
the new experimental error on $m_b$ ---\,obtained from the $n=2$ moment\,--- would be reduced to $13\,$MeV, making for a total uncertainty of $16\,$MeV if the theoretical error is estimated as in Ref.~\cite{Dehnadi:2015fra}.
%Since the largest uncertainty source is the experimental one, if it is drastically reduced the total uncertainty could go down by $\sim 40\%$. {\color{red}[Where does this number come from? I think we cannot make such a statement at all as it requires an assumption on how the new data looks like. I would like to delete the sentence completely.]} 
If the next perturbative order becomes available, the theoretical uncertainty may be reduced by $30\%$ independent of the way how the perturbative truncation error is estimated.

Concerning non-relativistic sum rules (for which non-perturbative effects are smaller than in quarkonium masses) we discuss the following two analyses:~Ref.~\cite{Beneke:2014pta} in which (potential) pNRQCD is used  (which provided N$^3$LO fixed-order non-relativistic calculations) together with the PS short-distance mass, and Ref.~\cite{Hoang:2012us} which uses (velocity) vNRQCD (which provided NNLL renormalization group improved non-relativistic calculations) and employs the 1S short-distance mass. The result from Ref.~\cite{Beneke:2014pta} is: 
\begin{equation}
  m_b(m_b)= 4.193_{-0.035}^{+0.022}\,\rm GeV,
\end{equation}
  and Ref. ~\cite{Hoang:2012us} yields:
\begin{equation}
  m_b(m_b)= 4.235 \pm 0.055\,\rm GeV.
\end{equation}
where the theoretical uncertainties are an order of magnitude larger than the experimental ones.
Improvements on this kind of bottom mass measurements can only come from the theoretical
side:~additional perturbative orders may become available (either through fixed-order
corrections or anomalous dimensions for the summation of logarithms) or an improved determination on the effects of the finite charm quark mass could be reached. In either case an error reduction will require a significant theoretical effort.
%Either way uncertainties should not drop down
%by more than $\sim 20\%$. {\color{red}[I do not understand the previous sentence. %How can we get to such a statement??]}

In the light of the available results as of 2022 one can envisage that the most precise determination in the future will come from relativistic sum rules
(although it is not inconceivable that on the time scale of some of the future collider facilities discussed in the Snowmass study lattice simulations with
dynamical bottom quarks are feasible). Assuming new data becomes available,
together with one extra perturbative coefficient, and hopefully new theoretical
strategies based on a better understanding of the perturbative series (see
e.g.~Ref.~\cite{Boito:2021wbj}, where optimized moments were tentatively
introduced), it is not unrealistic to assume that the uncertainty on $m_b(m_b)$ 
is reduced below $10\,$MeV. 

%\textbf{Some words to summarize and a quantitative estimate of the mb(mb) world average anno 2040?} Good enough? {\color{red}[Why is the year number 2040 coming up here? Do I miss something?]}

\section{The bottom-quark mass from three-jet rates}

Future $e^+e^-$ colliders can improve the precision of the $m_b(m_Z)$ measurement. A dedicated high-luminosity run at the $Z$-pole, i.e. the ``GigaZ" programme of a linear collider or the ``TeraZ'' run at the circular colliders, yields a sample of $Z$-bosons that exceeds that of the LEP experiments and SLD by orders of magnitude. We adopt the extrapolation of LEP/SLD results in Ref.~\cite{ILDnote2020} that assumes that the extraction of $m_b(m_Z)$ from the three-jet rates will be limited by the theory uncertainty and hadronization uncertainties. Both sources of uncertainty are assumed to be reduced by a factor 2. This requires fixed-order calculations at NNLO accuracy, with full consideration of mass effects, which is available for Higgs decays~\cite{Bernreuther:2018ynm}.  

The Higgs factory program itself, with several inverse attobarn at a center-of-mass energy of 240-250~\GeV{}, can take advantage of radiative-return events. The Lorentz-boost of the $Z$-bosons complicates the selection, reconstruction and interpretation. A dedicated full-simulation study is therefore required to provide a reliable, quantitative projection. However, it is clear that the radiative-return data has the potential to significantly improve the precision of existing LEP/SLC analysis.

Finally, a high-energy electron-positron collider operated at a center-of-mass energy of 250~\GeV{} or above can extend the analysis to higher energies and thus probe the effect of coloured states with masses heavier than that the Higgs boson on the running of the bottom quark mass. The potential of the three-jet rate measurement to determine $m_b(\mu)$ for $\mu=$ 250~\GeV{} has been studied in Ref.~\cite{ILDnote2020}. The mass dependence of the observable is found to drop rapidly with increasing $\mu$, since the bottom quark mass dependence is a power-suppressed correction. A measurement with a precision of 1~\GeV{} is feasible for $\mu=$ 250~\GeV{}.

\section{The bottom-quark mass from $Z-$boson decay}

The bottom quark mass at the scale of the $Z$-boson mass can also be inferred from the $Z \rightarrow b\bar{b}$ decay width. Currently, this method does not offer a competitive precision. Using $R_{0,b} = \Gamma(Z \rightarrow b\bar{b})/\Gamma_{\rm total} = 0.21582 \pm 0.00066$, as reported by the LEP/SLC Electro-weak Working Group~\cite{ALEPH:2005ab}, Ref.~\cite{Kluth:2022ucw} finds an uncertainty greater than 1~\GeV.

A future high-statistics $Z$-pole run, together with theory improvements, can significantly enhance the potential of this approach. Following the FCCee Conceptual Design Report~\cite{FCC:2018byv,FCC:2018evy}, that predicts a ten-fold increase of the precision of $R_{0,b}$,  one can expect a precision of 140~MeV (5\%) on $m_b(m_Z)$ after the ``TeraZ'' program.
This requires considerable improvements in the modelling of B- and D-hadron decays, compared to the reference analysis performed by SLC that forms the basis for the extrapolation by the FCCee study.

\section{The bottom-quark mass from Higgs decay}

The measurement of $m_b(m_H)$ from the Higgs decay width to a bottom-antibottom quark pair is expected to increase rapidly in precision as the precision of Higgs coupling measurement improves.
The method of Ref.~\cite{Aparisi:2021tym} provides a very clean theoretical basis that allows for steady progress as the experimental precision improves. The key aspect of this method is that the Higgs boson is a color-less spin-0 state with a relatively small decay width, such that the analysis is essentially insensitive to the theoretical knowledge of the Higgs production rate. For the same reason very precise theoretical predictions can be made for the Higgs partial width into a bottom-antibottom quark pair.

At the relevant dynamical scale, $m_H$, the QCD corrections are very well under control using $\mu\sim m_H$ as the renormalization scales of $\alpha_s$ and $m_b$. The partial width $\Gamma(H\rightarrow b\bar{b}$ is proportional to the squared of the bottom quark mass. This dependence arises because the decay is governed the bottom Yukawa coupling. For these reasons we expect the determination of $m_b(m_H)$ in $H\rightarrow b\bar{b}$ decay to become the ``golden'' measurement among the high-energy determinations. 

The current theory uncertainty from missing higher orders and parametric uncertainties from $\alpha_s$ and $m_H$ is estimated to be 60~\MeV{}~\cite{Aparisi:2021tym}, well below the current experimental precision. The theory uncertainty is dominated by the parametric uncertainty from the Higgs boson mass. The current uncertainty on the Higgs mass of 240~\MeV{} leads to an uncertainty of $\sim$40~MeV on $m_b(m_H)$ and is expected to come down considerably as more precise determinations of $m_H$ appear. Future prospects for Higgs mass measurements are summarized in Ref.~\cite{deBlas:2019rxi}. Both the HL-LHC~\cite{Cepeda:2019klc} and the Higgs factory~\cite{Yan:2016xyx} are expected to provide a measurement of the Higgs boson mass to 10-20~\MeV{} precision, which is sufficient to reduce the impact of this source of uncertainty on $m_b(m_H)$ to below 10~\MeV{}. 

The effect of the strong coupling $\alpha_s$, which amounts to an 0.2\% uncertainty in the ratio of branching ratios for an uncertainty of $0.001$ in the value of $\alpha_s(m_Z)$, is relatively small. A much larger parametric $\alpha_s$ uncertainty was reported in Ref.~\cite{deFlorian:2016spz}, which stems mainly from the parametric dependence of the evolution of the bottom quark mass from $\mu = m_b$ to $\mu = m_H$ on the value of $\alpha_s$. This source of uncertainty does not exist in the measurement of $m_b(m_H)$. The parametric uncertainty from the value of $\alpha_s$ is expected to remain sub-dominant even with only a very modest improvement of its world average. 

The dominant theory uncertainty is due to missing higher order electroweak corrections for the branching ratios, that are currently known with NLO precision. These electroweak uncertainties are about a factor of two larger than the current uncertainties from QCD. Thus the knowledge of the leading NNLO EW correction is expected to be sufficient to take full advantage of the power of the Higgs factory data. 
%The ATLAS prospect is 30-50 MeV for ZZ (ATL-PHYS-PUB-2018-054), while the more recent study in Ref.~\cite{Cepeda:2019klc} expects 10-20~\MeV{} for HL-LHC~\cite{deBlas:2019rxi}. possibly less than 10~\MeV{} for the Higgs factories (ILC: ~\cite{Yan:2016xyx}).

Ref.~\cite{Cepeda:2019klc} provides the projections for the LHC and its luminosity upgrade, extrapolating the partial run 2 results under the following assumptions: both statistical and systematic uncertainties are envisaged to scale with integrated luminosity $L$ as $1/\sqrt{L}$ up to certain limits, while theory uncertainties are expected to improve by a factor two. This ``S2 scenario'' leads to a projected uncertainty on the Higgs branching ratio to bottom quarks of 4.4\% (1.5\% stat., 1.3\% exp., 4.0\% theo.) and on $\lambda_{bz} = \mu^{bb}/\mu^{ZZ}$ of 3.1\% (1.3\% stat., 1.3\% syst., 2.6\% theo.), an improvement by nearly a factor of ten with respect to the first measurement in Ref.~\cite{Aparisi:2021tym}. %{\color{red}[I do not understand which analysis is referred to in the previous sentence. Is it Ref.~\cite{Aparisi:2021tym}? I may be better to add an explicit reference.]} The progress will lead to a rapid increase in the precision of $m_b(m_H)$, reaching 60~\MeV{} at the end of the HL-LHC program. 

%If this scenario is indeed realized and measurements continue to agree with the SM, this determination of $m_b(m_H)$  becomes the ideal probe for the scale evolution of $m_b$.

%deBlas:2019rxi

The Higgs boson couplings will be measured to even higher precision at future $e^+e^-$ Higgs factories~\cite{deBlas:2019rxi}. The recoil analysis yields a precisely measured production cross section, and normalization of the total width, while the branching fractions can be inferred from a simple counting experiment on the selected events. Sub-\% precision is expected for the $Hb\bar{b}$ coupling, based on detailed prospect studies of Ref.~\cite{Barklow:2017suo,Abramowicz:2016zbo}. 

For a numerical estimate of the potential of the Higgs factory projects to improve $m_b(m_H)$ we adopt the projections for the ratio $BR(H \rightarrow b\bar{b}) / BR(H \rightarrow W^+W^-$ from the ILC project. This ratio can be more precisely measured than the ratio $ BR (H \rightarrow b\bar{b} ) / BR(H \rightarrow ZZ) $, due to the better statistical precision in the $H\rightarrow W^+W^-$ channel. The 250~\gev{} stage of the International Linear Collider (ILC) can measure the former ratio to 0.86\% precision~\cite{Fujii:2019zll,Bambade:2019fyw}. The corresponding uncertainty on $m_b(m_H)$ is $\pm$12~\MeV. The complete ILC programme with stages at 250~\GeV{} and 500~\GeV{} is expected to improve the precision further, to 0.47\% on $BR(H\rightarrow b\bar{b})/BR(H\rightarrow W^+W^-)$, and yields an experimental uncertainty of 6~\MeV{} on $m_b(m_H)$.  Other Higgs factory projects are expected to reach similar precision.

At this stage, the experimental precision of $m_b(m_H)$ is expected to reach a relative precision of 0.2\%, a factor two better than the current world average for $m_b(m_b)$. At this point, additional theoretical care will be needed to assess the uncertainties in the theoretical predictions that are used for extraction of $m_b(m_H)$ and in relating it to $m_b(m_b)$. 

\section{The strong coupling}

A thorough and precise test of the scale evolution of the bottom quark mass requires precise values for the strong coupling, both at the scale of the bottom quark mass and at the electro-weak scale $\mu \sim m_Z \sim m_H$. In the traditional approach to the determination of quark masses, one usually assumes that the strong coupling is an external parameter that leads only to parametric uncertainties. In this context, it is deemed sufficient to consider the PDG world average for $\alpha_s(m_Z)$ (or at any other renormalization scale $\mu$) and use the Standard Model evolution equation to determine the strong coupling at the scale needed for the theoretical calculations. This is a reasonable approach for the determination of the bottom quark mass at a given scale, but for a precise test of the QCD evolution one must apply a more conservative view. The running of the strong coupling may be affected by the same new physics effects that alter the running of the quark masses and therefore cannot be assumed to follow the Standard Model RGE evolution. The analysis of the scale evolution of the bottom mass for scales between $m_b$ and $m_Z$ or $m_H$ therefore also requires precise measurements of the strong coupling $\alpha_s(\mu)$ over the interval of scales considered. 

Lattice determinations of the strong coupling have achieved sub-\% precision ~\cite{Aoki:2021kgd}. The Flavour Lattice Averaging Group (FLAG) working group on the strong coupling constant, provides the following projection in Ref.~\cite{alphas2022}: ``a total error clearly below half a percent for $\alpha_s(m_Z)$ seems achievable within the next few years by pushing the step-scaling method further, possibly in combination with the decoupling strategy." 

%We adopt this statement as our projection for the low-energy determination of the strong coupling constant, assigning a relative uncertainty of 0.5\% to $\alpha_s(m_b)$. 
% 0.0006
%Measurements from  $\tau$-lepton~\cite{Pich:2020qna} decays and quarkonium sum rules..... Progress is expected ....... 

An independent low-energy determination of the strong coupling comes from determinations from $\tau$-decays~\cite{alphas2022}. Today, these yield a value of $\alpha_s(m_\tau) = 0.3077 \pm 0.0065 (exp.) \pm 0.0038 (theory)$~\cite{Boito:2020xli}. The dominant experimental uncertainties are expected to be reduced strongly using Belle II data~\cite{Belle:2008xpe} and eventually new $Z$-pole data at a future electron-positron collider~\cite{Dam:2021ibi}, potentially achieving sub-\% precision.

The determination of the strong coupling at the electro-weak scale is currently dominated by LEP measurements. In Ref.~\cite{Aparisi:2021tym} an uncertainty of 0.004 was assigned to $\alpha_s(m_Z)$ based solely on electroweak-scale measurements. A future electron-positron collider running at the $Z$ pole provides an ideal environment to reduce this uncertainty. Ref.~\cite{dEnterria:2020cpv} claims the strong coupling can be measured to better than 0.1\% exploiting $Z$-boson hadronic pseudo-observables, provided the theoretical uncertainties are reduced by incorporating missing higher-order QCD and mixed QCD+EW corrections~\cite{freitas2019theoretical}.

Based on these qualititative prospects, we assign an uncertainty of 0.5\% to the strong coupling over the range from $m_b$ to $m_H$.

\section{Projection for the test of the scale evolution}

\begin{figure}[h!]
\includegraphics[width=0.99\columnwidth]{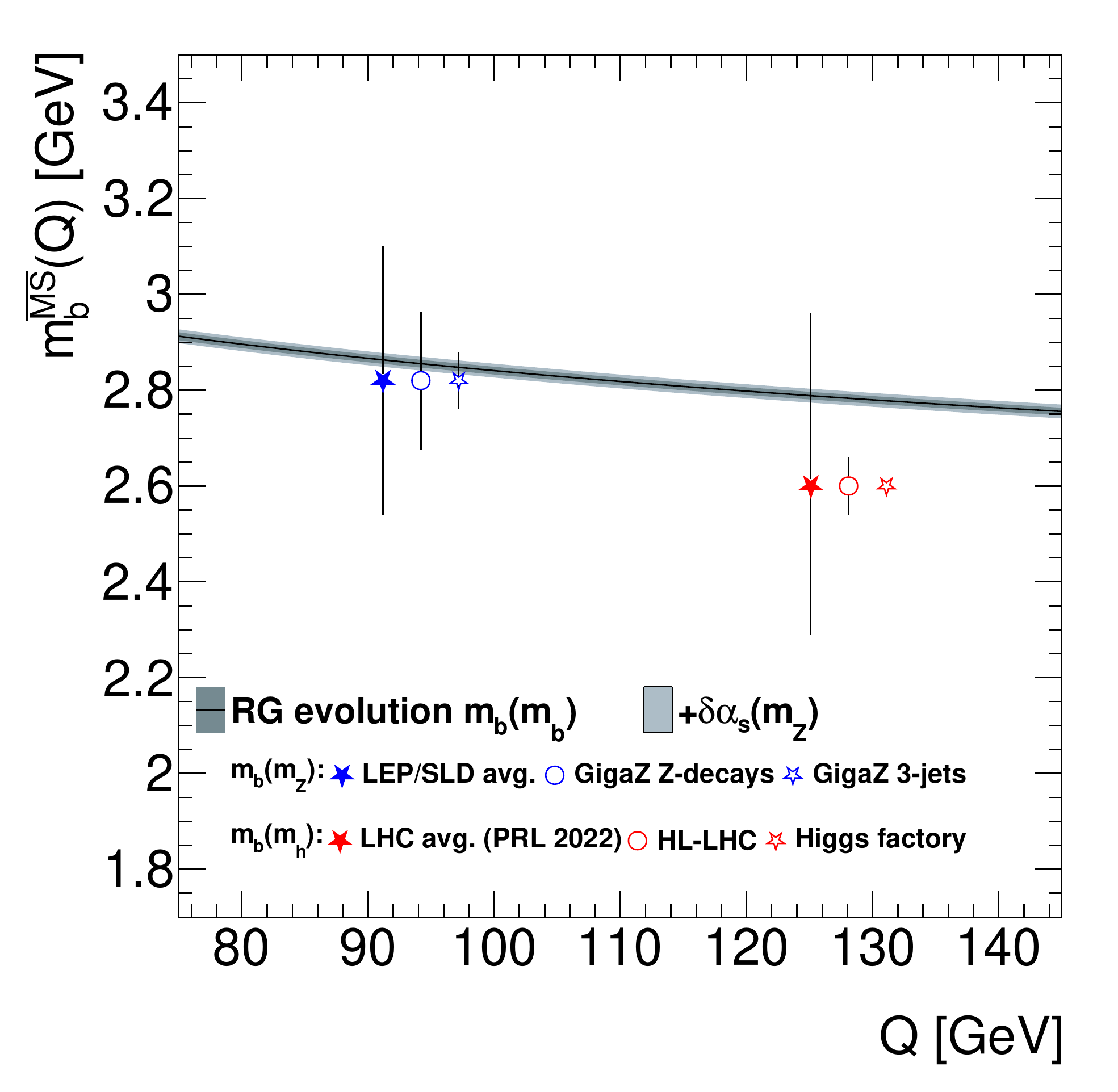}%
\caption{\label{fig:running_mass_projection} Prospects for measurements of the scale evolution of the bottom quark \MSbar{} mass at future colliders. The markers are projections for $m_b(m_Z)$ from three-jet rates at the $Z$-pole and for $m_b(m_H)$ from Higgs boson branching fractions. The RGE evolution of the mass is calculated at five-loop precision with REvolver~\cite{Hoang:2021fhn}. }
\end{figure}

The projections and extrapolations discussed in the previous sections have been included in Fig.~\ref{fig:running_mass_projection}. The markers are centered on the current central values for $m_b(m_Z)$ and $m_b(m_H)$ and the error bars indicate the projected precision. The solid line indicates the evolution of the PDG world average from $m_b(m_b)$ to a higher scale using the RGE calculation included in the REvolver code~\cite{Hoang:2021fhn} at five-loop precision. The uncertainty band includes the projected uncertainty of 10~\MeV{} on $m_b(m_b)$ (dark grey) and an 0.5\% uncertainty on $\alpha_s(m_Z)$. 
 
The independent determinations of the bottom quark mass at different energies yield a precision test of the scale evolution of the bottom quark mass. High-scale determinations can be used to search for the impact of new massive coloured states on the scale evolution, using a similar strategy to studies of $\alpha_s$~\cite{Llorente:2018wup,Jezabek:1992sq}, and possibly incorporating the analysis of $\alpha_s$ and $m_b$ in a combined fit. The implementation of this programme, and a precise estimate of its sensitivity, is left for future work.

\section{Summary}

In the next decades, with the completion of the high luminosity programme of the LHC and the construction of a new ``Higgs factory" electron-positron collider, rapid progress is envisaged in the measurement of Higgs coupling measurement. These precise measurements will enable an extraction of the \MSbar{} bottom quark mass $m_b(\mu)$ at the scale given by the Higgs boson mass, $m_b(m_H)$, with a precision of the order of 10~\MeV. With a relative precision of 2 per mille, the high-scale measurement can reach a similar precision as $m_b(m_b)$ based on low-energy measurements.

Together with improved measurements of $m_b(m_b)$ from low-energy data, $m_b(m_Z)$ from three-jet rates in $e^+e^-$ collisions (and possibly new measurements at scales smaller than $m_Z$ and larger than $m_H$), one can expect to map out the scale evolution of the bottom quark mass from $m_b$ to $m_H$ with a precision at the few per mille level. At the same time, improved measurements of the strong coupling at each of these scales reduce the uncertainty in the evolution between the two energies. When all these elements are brought together, they form a powerful test of the ``running" of quark masses predicted by the Standard Model and allow for stringent limits on coloured states with mass below the electroweak scale.

\begin{acknowledgments}
The authors than Stefan Sint for help understanding the lattice prospects and acknowledge support from projects FPA2015-65652-C4-3-R, PID2020-114473GB-I00, PID2019-105439GB-C22 and PGC2018-094856-B-100 (MICIN/AEI/10.13039/501100011033), support from the U. Valencia and CSIC for H. Yamamoto, PROMETEO/2021/071, PROMETEO-2018/060 and CIDEGENT/2020/21 (Generalitat Valenciana), iLINK grant LINKB20065 (CSIC), the FWF  Austrian  Science  Fund  Project  No.\ P28535-N27 and Doctoral Program No.\ W1252-N27; the EU STRONG-2020 project under the program H2020-INFRAIA-2018-1, grant agreement No.\ 824093 and the COST Action CA16201 PARTICLEFACE.
\end{acknowledgments}

\bibliography{bottom.bib}

%merlin.mbs apsrev4-1.bst 2010-07-25 4.21a (PWD, AO, DPC) hacked
%Control: key (0)
%Control: author (72) initials jnrlst
%Control: editor formatted (1) identically to author
%Control: production of article title (-1) disabled
%Control: page (0) single
%Control: year (1) truncated
%Control: production of eprint (0) enabled
\begin{thebibliography}{77}%
\makeatletter
\providecommand \@ifxundefined [1]{%
 \@ifx{#1\undefined}
}%
\providecommand \@ifnum [1]{%
 \ifnum #1\expandafter \@firstoftwo
 \else \expandafter \@secondoftwo
 \fi
}%
\providecommand \@ifx [1]{%
 \ifx #1\expandafter \@firstoftwo
 \else \expandafter \@secondoftwo
 \fi
}%
\providecommand \natexlab [1]{#1}%
\providecommand \enquote  [1]{``#1''}%
\providecommand \bibnamefont  [1]{#1}%
\providecommand \bibfnamefont [1]{#1}%
\providecommand \citenamefont [1]{#1}%
\providecommand \href@noop [0]{\@secondoftwo}%
\providecommand \href [0]{\begingroup \@sanitize@url \@href}%
\providecommand \@href[1]{\@@startlink{#1}\@@href}%
\providecommand \@@href[1]{\endgroup#1\@@endlink}%
\providecommand \@sanitize@url [0]{\catcode `\\12\catcode `\$12\catcode
  `\&12\catcode `\#12\catcode `\^12\catcode `\_12\catcode `\%12\relax}%
\providecommand \@@startlink[1]{}%
\providecommand \@@endlink[0]{}%
\providecommand \url  [0]{\begingroup\@sanitize@url \@url }%
\providecommand \@url [1]{\endgroup\@href {#1}{\urlprefix }}%
\providecommand \urlprefix  [0]{URL }%
\providecommand \Eprint [0]{\href }%
\providecommand \doibase [0]{http://dx.doi.org/}%
\providecommand \selectlanguage [0]{\@gobble}%
\providecommand \bibinfo  [0]{\@secondoftwo}%
\providecommand \bibfield  [0]{\@secondoftwo}%
\providecommand \translation [1]{[#1]}%
\providecommand \BibitemOpen [0]{}%
\providecommand \bibitemStop [0]{}%
\providecommand \bibitemNoStop [0]{.\EOS\space}%
\providecommand \EOS [0]{\spacefactor3000\relax}%
\providecommand \BibitemShut  [1]{\csname bibitem#1\endcsname}%
\let\auto@bib@innerbib\@empty
%</preamble>
\bibitem [{Note1()}]{Note1}%
  \BibitemOpen
  \bibinfo {note} {Throughout this White Paper we define the $\protect
  \overline {\protect \rm MS}$ bottom quark mass in the $n_f=5$ flavor
  scheme.}\BibitemShut {Stop}%
\bibitem [{\citenamefont {Vermaseren}\ \emph {et~al.}(1997)\citenamefont
  {Vermaseren}, \citenamefont {Larin},\ and\ \citenamefont {van
  Ritbergen}}]{Vermaseren:1997fq}%
  \BibitemOpen
  \bibfield  {author} {\bibinfo {author} {\bibfnamefont {J.~A.~M.}\
  \bibnamefont {Vermaseren}}, \bibinfo {author} {\bibfnamefont {S.~A.}\
  \bibnamefont {Larin}}, \ and\ \bibinfo {author} {\bibfnamefont
  {T.}~\bibnamefont {van Ritbergen}},\ }\href {\doibase
  10.1016/S0370-2693(97)00660-6} {\bibfield  {journal} {\bibinfo  {journal}
  {Phys. Lett. B}\ }\textbf {\bibinfo {volume} {405}},\ \bibinfo {pages} {327}
  (\bibinfo {year} {1997})},\ \Eprint {http://arxiv.org/abs/hep-ph/9703284}
  {arXiv:hep-ph/9703284} \BibitemShut {NoStop}%
\bibitem [{\citenamefont {Chetyrkin}(1997)}]{Chetyrkin:1997dh}%
  \BibitemOpen
  \bibfield  {author} {\bibinfo {author} {\bibfnamefont {K.~G.}\ \bibnamefont
  {Chetyrkin}},\ }\href {\doibase 10.1016/S0370-2693(97)00535-2} {\bibfield
  {journal} {\bibinfo  {journal} {Phys. Lett. B}\ }\textbf {\bibinfo {volume} {404}},\ \bibinfo {pages} {161} (\bibinfo {year} {1997})},\ \Eprint
  {http://arxiv.org/abs/hep-ph/9703278} {arXiv:hep-ph/9703278} \BibitemShut
  {NoStop}%
\bibitem [{\citenamefont {Baikov}\ \emph {et~al.}(2014)\citenamefont {Baikov},
  \citenamefont {Chetyrkin},\ and\ \citenamefont {K\"uhn}}]{Baikov:2014qja}%
  \BibitemOpen
  \bibfield  {author} {\bibinfo {author} {\bibfnamefont {P.~A.}\ \bibnamefont
  {Baikov}}, \bibinfo {author} {\bibfnamefont {K.~G.}\ \bibnamefont
  {Chetyrkin}}, \ and\ \bibinfo {author} {\bibfnamefont {J.~H.}\ \bibnamefont
  {K\"uhn}},\ }\href {\doibase 10.1007/JHEP10(2014)076} {\bibfield  {journal}
  {\bibinfo  {journal} {JHEP}\ }\textbf {\bibinfo {volume} {10}},\ \bibinfo
  {pages} {076} (\bibinfo {year} {2014})},\ \Eprint
  {http://arxiv.org/abs/1402.6611} {arXiv:1402.6611 [hep-ph]} \BibitemShut
  {NoStop}%
\bibitem [{\citenamefont {Herren}\ and\ \citenamefont
  {Steinhauser}(2018)}]{Herren:2017osy}%
  \BibitemOpen
  \bibfield  {author} {\bibinfo {author} {\bibfnamefont {F.}~\bibnamefont
  {Herren}}\ and\ \bibinfo {author} {\bibfnamefont {M.}~\bibnamefont
  {Steinhauser}},\ }\href {\doibase 10.1016/j.cpc.2017.11.014} {\bibfield
  {journal} {\bibinfo  {journal} {Comput. Phys. Commun.}\ }\textbf {\bibinfo
  {volume} {224}},\ \bibinfo {pages} {333} (\bibinfo {year} {2018})},\ \Eprint
  {http://arxiv.org/abs/1703.03751} {arXiv:1703.03751 [hep-ph]} \BibitemShut
  {NoStop}%
\bibitem [{\citenamefont {Hoang}\ \emph {et~al.}(2021)\citenamefont {Hoang},
  \citenamefont {Lepenik},\ and\ \citenamefont {Mateu}}]{Hoang:2021fhn}%
  \BibitemOpen
  \bibfield  {author} {\bibinfo {author} {\bibfnamefont {A.~H.}\ \bibnamefont
  {Hoang}}, \bibinfo {author} {\bibfnamefont {C.}~\bibnamefont {Lepenik}}, \
  and\ \bibinfo {author} {\bibfnamefont {V.}~\bibnamefont {Mateu}},\
  }\href@noop {} {\  (\bibinfo {year} {2021})},\ \Eprint
  {http://arxiv.org/abs/2102.01085} {arXiv:2102.01085 [hep-ph]} \BibitemShut
  {NoStop}%
\bibitem [{\citenamefont {Zyla}\ \emph {et~al.}(2020)\citenamefont {Zyla} \emph
  {et~al.}}]{Zyla:2020zbs}%
  \BibitemOpen
  \bibfield  {author} {\bibinfo {author} {\bibfnamefont {P.}~\bibnamefont
  {Zyla}} \emph {et~al.} (\bibinfo {collaboration} {Particle Data Group}),\
  }\href {\doibase 10.1093/ptep/ptaa104} {\bibfield  {journal} {\bibinfo
  {journal} {PTEP}\ }\textbf {\bibinfo {volume} {2020}},\ \bibinfo {pages}
  {083C01} (\bibinfo {year} {2020})}\BibitemShut {NoStop}%
\bibitem [{\citenamefont {Gizhko}\ \emph {et~al.}(2017)\citenamefont {Gizhko}
  \emph {et~al.}}]{Gizhko:2017fiu}%
  \BibitemOpen
  \bibfield  {author} {\bibinfo {author} {\bibfnamefont {A.}~\bibnamefont
  {Gizhko}} \emph {et~al.},\ }\href {\doibase 10.1016/j.physletb.2017.11.002}
  {\bibfield  {journal} {\bibinfo  {journal} {Phys. Lett.}\ }\textbf {\bibinfo
  {volume} {B775}},\ \bibinfo {pages} {233} (\bibinfo {year} {2017})},\ \Eprint
  {http://arxiv.org/abs/1705.08863} {arXiv:1705.08863 [hep-ph]} \BibitemShut
  {NoStop}%
%%CITATION = ARXIV:1705.08863;%%
\bibitem [{\citenamefont {{CMS collaboration}}(2020)}]{Sirunyan:2019jyn}%
  \BibitemOpen
  \bibfield  {author} {\bibinfo {author} {\bibnamefont {{CMS collaboration}}},\
  }\href {\doibase 10.1016/j.physletb.2020.135263} {\bibfield  {journal}
  {\bibinfo  {journal} {Phys. Lett. B}\ }\textbf {\bibinfo {volume} {803}},\
  \bibinfo {pages} {135263} (\bibinfo {year} {2020})},\ \Eprint
  {http://arxiv.org/abs/1909.09193} {arXiv:1909.09193 [hep-ex]} \BibitemShut
  {NoStop}%
\bibitem [{\citenamefont {Narison}(2020)}]{Narison:2019tym}%
  \BibitemOpen
  \bibfield  {author} {\bibinfo {author} {\bibfnamefont {S.}~\bibnamefont
  {Narison}},\ }\href {\doibase 10.1016/j.physletb.2020.135221} {\bibfield
  {journal} {\bibinfo  {journal} {Phys. Lett. B}\ }\textbf {\bibinfo {volume}
  {802}},\ \bibinfo {pages} {135221} (\bibinfo {year} {2020})},\ \Eprint
  {http://arxiv.org/abs/1906.03614} {arXiv:1906.03614 [hep-ph]} \BibitemShut
  {NoStop}%
\bibitem [{\citenamefont {Peset}\ \emph {et~al.}(2018)\citenamefont {Peset},
  \citenamefont {Pineda},\ and\ \citenamefont {Segovia}}]{Peset:2018ria}%
  \BibitemOpen
  \bibfield  {author} {\bibinfo {author} {\bibfnamefont {C.}~\bibnamefont
  {Peset}}, \bibinfo {author} {\bibfnamefont {A.}~\bibnamefont {Pineda}}, \
  and\ \bibinfo {author} {\bibfnamefont {J.}~\bibnamefont {Segovia}},\ }\href
  {\doibase 10.1007/JHEP09(2018)167} {\bibfield  {journal} {\bibinfo  {journal}
  {JHEP}\ } {\bibinfo {volume} {09}},\ \bibinfo {pages} {167} (\bibinfo
  {year} {2018})},\ \Eprint {http://arxiv.org/abs/1806.05197} {arXiv:1806.05197
  [hep-ph]} \BibitemShut {NoStop}%
\bibitem [{\citenamefont {Kiyo}\ \emph {et~al.}(2016)\citenamefont {Kiyo},
  \citenamefont {Mishima},\ and\ \citenamefont {Sumino}}]{Kiyo:2015ufa}%
  \BibitemOpen
  \bibfield  {author} {\bibinfo {author} {\bibfnamefont {Y.}~\bibnamefont
  {Kiyo}}, \bibinfo {author} {\bibfnamefont {G.}~\bibnamefont {Mishima}}, \
  and\ \bibinfo {author} {\bibfnamefont {Y.}~\bibnamefont {Sumino}},\ }\href
  {\doibase 10.1016/j.physletb.2015.11.040, 10.1016/j.physletb.2017.09.024}
  {\bibfield  {journal} {\bibinfo  {journal} {Phys. Lett.}\ }\textbf {\bibinfo
  {volume} {B752}},\ \bibinfo {pages} {122} (\bibinfo {year} {2016})},\
  \bibinfo {note} {[Erratum: Phys. Lett.B772,878(2017)]},\ \Eprint
  {http://arxiv.org/abs/1510.07072} {arXiv:1510.07072 [hep-ph]} \BibitemShut
  {NoStop}%
%%CITATION = ARXIV:1510.07072;%%
\bibitem [{\citenamefont {Penin}\ and\ \citenamefont
  {Zerf}(2014)}]{Penin:2014zaa}%
  \BibitemOpen
  \bibfield  {author} {\bibinfo {author} {\bibfnamefont {A.~A.}\ \bibnamefont
  {Penin}}\ and\ \bibinfo {author} {\bibfnamefont {N.}~\bibnamefont {Zerf}},\
  }\href {\doibase 10.1007/JHEP04(2014)120} {\bibfield  {journal} {\bibinfo
  {journal} {JHEP}\ }\textbf {\bibinfo {volume} {04}},\ \bibinfo {pages} {120}
  (\bibinfo {year} {2014})},\ \Eprint {http://arxiv.org/abs/1401.7035}
  {arXiv:1401.7035 [hep-ph]} \BibitemShut {NoStop}%
%%CITATION = ARXIV:1401.7035;%%
\bibitem [{\citenamefont {Alberti}\ \emph {et~al.}(2015)\citenamefont
  {Alberti}, \citenamefont {Gambino}, \citenamefont {Healey},\ and\
  \citenamefont {Nandi}}]{Alberti:2014yda}%
  \BibitemOpen
  \bibfield  {author} {\bibinfo {author} {\bibfnamefont {A.}~\bibnamefont
  {Alberti}}, \bibinfo {author} {\bibfnamefont {P.}~\bibnamefont {Gambino}},
  \bibinfo {author} {\bibfnamefont {K.~J.}\ \bibnamefont {Healey}}, \ and\
  \bibinfo {author} {\bibfnamefont {S.}~\bibnamefont {Nandi}},\ }\href
  {\doibase 10.1103/PhysRevLett.114.061802} {\bibfield  {journal} {\bibinfo
  {journal} {Phys. Rev. Lett.}\ }\textbf {\bibinfo {volume} {114}},\ \bibinfo
  {pages} {061802} (\bibinfo {year} {2015})},\ \Eprint
  {http://arxiv.org/abs/1411.6560} {arXiv:1411.6560 [hep-ph]} \BibitemShut
  {NoStop}%
\bibitem [{\citenamefont {Beneke}\ \emph {et~al.}(2015)\citenamefont {Beneke},
  \citenamefont {Maier}, \citenamefont {Piclum},\ and\ \citenamefont
  {Rauh}}]{Beneke:2014pta}%
  \BibitemOpen
  \bibfield  {author} {\bibinfo {author} {\bibfnamefont {M.}~\bibnamefont
  {Beneke}}, \bibinfo {author} {\bibfnamefont {A.}~\bibnamefont {Maier}},
  \bibinfo {author} {\bibfnamefont {J.}~\bibnamefont {Piclum}}, \ and\ \bibinfo
  {author} {\bibfnamefont {T.}~\bibnamefont {Rauh}},\ }\href {\doibase
  10.1016/j.nuclphysb.2014.12.001} {\bibfield  {journal} {\bibinfo  {journal}
  {Nucl. Phys. B}\ } {\bibinfo {volume} {891}},\ \bibinfo {pages} {42}
  (\bibinfo {year} {2015})},\ \Eprint {http://arxiv.org/abs/1411.3132}
  {arXiv:1411.3132 [hep-ph]} \BibitemShut {NoStop}%
\bibitem [{\citenamefont {Dehnadi}\ \emph {et~al.}(2015)\citenamefont
  {Dehnadi}, \citenamefont {Hoang},\ and\ \citenamefont
  {Mateu}}]{Dehnadi:2015fra}%
  \BibitemOpen
  \bibfield  {author} {\bibinfo {author} {\bibfnamefont {B.}~\bibnamefont
  {Dehnadi}}, \bibinfo {author} {\bibfnamefont {A.~H.}\ \bibnamefont {Hoang}},
  \ and\ \bibinfo {author} {\bibfnamefont {V.}~\bibnamefont {Mateu}},\ }\href
  {\doibase 10.1007/JHEP08(2015)155} {\bibfield  {journal} {\bibinfo  {journal}
  {JHEP}\ } {\bibinfo {volume} {08}},\ \bibinfo {pages} {155} (\bibinfo
  {year} {2015})},\ \Eprint {http://arxiv.org/abs/1504.07638} {arXiv:1504.07638
  [hep-ph]} \BibitemShut {NoStop}%
\bibitem [{\citenamefont {Lucha}\ \emph {et~al.}(2013)\citenamefont {Lucha},
  \citenamefont {Melikhov},\ and\ \citenamefont {Simula}}]{Lucha:2013gta}%
  \BibitemOpen
  \bibfield  {author} {\bibinfo {author} {\bibfnamefont {W.}~\bibnamefont
  {Lucha}}, \bibinfo {author} {\bibfnamefont {D.}~\bibnamefont {Melikhov}}, \
  and\ \bibinfo {author} {\bibfnamefont {S.}~\bibnamefont {Simula}},\ }\href
  {\doibase 10.1103/PhysRevD.88.056011} {\bibfield  {journal} {\bibinfo
  {journal} {Phys. Rev. D}\ }\textbf {\bibinfo {volume} {88}},\ \bibinfo
  {pages} {056011} (\bibinfo {year} {2013})},\ \Eprint
  {http://arxiv.org/abs/1305.7099} {arXiv:1305.7099 [hep-ph]} \BibitemShut
  {NoStop}%
\bibitem [{\citenamefont {Bodenstein}\ \emph {et~al.}(2012)\citenamefont
  {Bodenstein}, \citenamefont {Bordes}, \citenamefont {Dominguez},
  \citenamefont {Penarrocha},\ and\ \citenamefont
  {Schilcher}}]{Bodenstein:2011fv}%
  \BibitemOpen
  \bibfield  {author} {\bibinfo {author} {\bibfnamefont {S.}~\bibnamefont
  {Bodenstein}}, \bibinfo {author} {\bibfnamefont {J.}~\bibnamefont {Bordes}},
  \bibinfo {author} {\bibfnamefont {C.}~\bibnamefont {Dominguez}}, \bibinfo
  {author} {\bibfnamefont {J.}~\bibnamefont {Penarrocha}}, \ and\ \bibinfo
  {author} {\bibfnamefont {K.}~\bibnamefont {Schilcher}},\ }\href {\doibase
  10.1103/PhysRevD.85.034003} {\bibfield  {journal} {\bibinfo  {journal} {Phys.
  Rev. D}\ }\textbf {\bibinfo {volume} {85}},\ \bibinfo {pages} {034003}
  (\bibinfo {year} {2012})},\ \Eprint {http://arxiv.org/abs/1111.5742}
  {arXiv:1111.5742 [hep-ph]} \BibitemShut {NoStop}%
\bibitem [{\citenamefont {Laschka}\ \emph {et~al.}(2011)\citenamefont
  {Laschka}, \citenamefont {Kaiser},\ and\ \citenamefont
  {Weise}}]{Laschka:2011zr}%
  \BibitemOpen
  \bibfield  {author} {\bibinfo {author} {\bibfnamefont {A.}~\bibnamefont
  {Laschka}}, \bibinfo {author} {\bibfnamefont {N.}~\bibnamefont {Kaiser}}, \
  and\ \bibinfo {author} {\bibfnamefont {W.}~\bibnamefont {Weise}},\ }\href
  {\doibase 10.1103/PhysRevD.83.094002} {\bibfield  {journal} {\bibinfo
  {journal} {Phys. Rev. D}\ }\textbf {\bibinfo {volume} {83}},\ \bibinfo
  {pages} {094002} (\bibinfo {year} {2011})},\ \Eprint
  {http://arxiv.org/abs/1102.0945} {arXiv:1102.0945 [hep-ph]} \BibitemShut
  {NoStop}%
\bibitem [{\citenamefont {Chetyrkin}\ \emph {et~al.}(2009)\citenamefont
  {Chetyrkin}, \citenamefont {Kuhn}, \citenamefont {Maier}, \citenamefont
  {Maierhofer}, \citenamefont {Marquard}, \citenamefont {Steinhauser},\ and\
  \citenamefont {Sturm}}]{Chetyrkin:2009fv}%
  \BibitemOpen
  \bibfield  {author} {\bibinfo {author} {\bibfnamefont {K.}~\bibnamefont
  {Chetyrkin}}, \bibinfo {author} {\bibfnamefont {J.}~\bibnamefont {Kuhn}},
  \bibinfo {author} {\bibfnamefont {A.}~\bibnamefont {Maier}}, \bibinfo
  {author} {\bibfnamefont {P.}~\bibnamefont {Maierhofer}}, \bibinfo {author}
  {\bibfnamefont {P.}~\bibnamefont {Marquard}}, \bibinfo {author}
  {\bibfnamefont {M.}~\bibnamefont {Steinhauser}}, \ and\ \bibinfo {author}
  {\bibfnamefont {C.}~\bibnamefont {Sturm}},\ }\href {\doibase
  10.1103/PhysRevD.80.074010} {\bibfield  {journal} {\bibinfo  {journal} {Phys.
  Rev. D}\ } {\bibinfo {volume} {80}},\ \bibinfo {pages} {074010}
  (\bibinfo {year} {2009})},\ \Eprint {http://arxiv.org/abs/0907.2110}
  {arXiv:0907.2110 [hep-ph]} \BibitemShut {NoStop}%
\bibitem [{\citenamefont {Bazavov}\ \emph {et~al.}(2018)\citenamefont {Bazavov}
  \emph {et~al.}}]{Bazavov:2018omf}%
  \BibitemOpen
  \bibfield  {author} {\bibinfo {author} {\bibfnamefont {A.}~\bibnamefont
  {Bazavov}} \emph {et~al.} (\bibinfo {collaboration} {Fermilab Lattice, MILC,
  TUMQCD}),\ }\href {\doibase 10.1103/PhysRevD.98.054517} {\bibfield  {journal}
  {\bibinfo  {journal} {Phys. Rev. D}\ }\textbf {\bibinfo {volume} {98}},\
  \bibinfo {pages} {054517} (\bibinfo {year} {2018})},\ \Eprint
  {http://arxiv.org/abs/1802.04248} {arXiv:1802.04248 [hep-lat]} \BibitemShut
  {NoStop}%
\bibitem [{\citenamefont {Colquhoun}\ \emph {et~al.}(2015)\citenamefont
  {Colquhoun}, \citenamefont {Dowdall}, \citenamefont {Davies}, \citenamefont
  {Hornbostel},\ and\ \citenamefont {Lepage}}]{Colquhoun:2014ica}%
  \BibitemOpen
  \bibfield  {author} {\bibinfo {author} {\bibfnamefont {B.}~\bibnamefont
  {Colquhoun}}, \bibinfo {author} {\bibfnamefont {R.}~\bibnamefont {Dowdall}},
  \bibinfo {author} {\bibfnamefont {C.}~\bibnamefont {Davies}}, \bibinfo
  {author} {\bibfnamefont {K.}~\bibnamefont {Hornbostel}}, \ and\ \bibinfo
  {author} {\bibfnamefont {G.}~\bibnamefont {Lepage}},\ }\href {\doibase
  10.1103/PhysRevD.91.074514} {\bibfield  {journal} {\bibinfo  {journal} {Phys.
  Rev. D}\ }\textbf {\bibinfo {volume} {91}},\ \bibinfo {pages} {074514}
  (\bibinfo {year} {2015})},\ \Eprint {http://arxiv.org/abs/1408.5768}
  {arXiv:1408.5768 [hep-lat]} \BibitemShut {NoStop}%
\bibitem [{\citenamefont {Bernardoni}\ \emph {et~al.}(2014)\citenamefont
  {Bernardoni} \emph {et~al.}}]{Bernardoni:2013xba}%
  \BibitemOpen
  \bibfield  {author} {\bibinfo {author} {\bibfnamefont {F.}~\bibnamefont
  {Bernardoni}} \emph {et~al.},\ }\href {\doibase
  10.1016/j.physletb.2014.01.046} {\bibfield  {journal} {\bibinfo  {journal}
  {Phys. Lett. B}\ }\textbf {\bibinfo {volume} {730}},\ \bibinfo {pages} {171}
  (\bibinfo {year} {2014})},\ \Eprint {http://arxiv.org/abs/1311.5498}
  {arXiv:1311.5498 [hep-lat]} \BibitemShut {NoStop}%
\bibitem [{\citenamefont {Lee}\ \emph {et~al.}(2013)\citenamefont {Lee},
  \citenamefont {Monahan}, \citenamefont {Horgan}, \citenamefont {Davies},
  \citenamefont {Dowdall},\ and\ \citenamefont {Koponen}}]{Lee:2013mla}%
  \BibitemOpen
  \bibfield  {author} {\bibinfo {author} {\bibfnamefont {A.}~\bibnamefont
  {Lee}}, \bibinfo {author} {\bibfnamefont {C.}~\bibnamefont {Monahan}},
  \bibinfo {author} {\bibfnamefont {R.}~\bibnamefont {Horgan}}, \bibinfo
  {author} {\bibfnamefont {C.}~\bibnamefont {Davies}}, \bibinfo {author}
  {\bibfnamefont {R.}~\bibnamefont {Dowdall}}, \ and\ \bibinfo {author}
  {\bibfnamefont {J.}~\bibnamefont {Koponen}} (\bibinfo {collaboration}
  {HPQCD}),\ }\href {\doibase 10.1103/PhysRevD.87.074018} {\bibfield  {journal}
  {\bibinfo  {journal} {Phys. Rev. D}\ }\textbf {\bibinfo {volume} {87}},\
  \bibinfo {pages} {074018} (\bibinfo {year} {2013})},\ \Eprint
  {http://arxiv.org/abs/1302.3739} {arXiv:1302.3739 [hep-lat]} \BibitemShut
  {NoStop}%
\bibitem [{\citenamefont {Dimopoulos}\ \emph {et~al.}(2012)\citenamefont
  {Dimopoulos} \emph {et~al.}}]{Dimopoulos:2011gx}%
  \BibitemOpen
  \bibfield  {author} {\bibinfo {author} {\bibfnamefont {P.}~\bibnamefont
  {Dimopoulos}} \emph {et~al.} (\bibinfo {collaboration} {ETM}),\ }\href
  {\doibase 10.1007/JHEP01(2012)046} {\bibfield  {journal} {\bibinfo  {journal}
  {JHEP}\ }\textbf {\bibinfo {volume} {01}},\ \bibinfo {pages} {046} (\bibinfo
  {year} {2012})},\ \Eprint {http://arxiv.org/abs/1107.1441} {arXiv:1107.1441
  [hep-lat]} \BibitemShut {NoStop}%
\bibitem [{\citenamefont {Aoki}\ \emph {et~al.}(2020)\citenamefont {Aoki} \emph
  {et~al.}}]{Aoki:2019cca}%
  \BibitemOpen
  \bibfield  {author} {\bibinfo {author} {\bibfnamefont {S.}~\bibnamefont
  {Aoki}} \emph {et~al.} (\bibinfo {collaboration} {Flavour Lattice Averaging
  Group}),\ }\href {\doibase 10.1140/epjc/s10052-019-7354-7} {\bibfield
  {journal} {\bibinfo  {journal} {Eur. Phys. J. C}\ }\textbf {\bibinfo {volume}
  {80}},\ \bibinfo {pages} {113} (\bibinfo {year} {2020})},\ \Eprint
  {http://arxiv.org/abs/1902.08191} {arXiv:1902.08191 [hep-lat]} \BibitemShut
  {NoStop}%
\bibitem [{\citenamefont {Abramowicz}\ \emph {et~al.}(2018)\citenamefont
  {Abramowicz} \emph {et~al.}}]{H1:2018flt}%
  \BibitemOpen
  \bibfield  {author} {\bibinfo {author} {\bibfnamefont {H.}~\bibnamefont
  {Abramowicz}} \emph {et~al.} (\bibinfo {collaboration} {H1, ZEUS}),\ }\href
  {\doibase 10.1140/epjc/s10052-018-5848-3} {\bibfield  {journal} {\bibinfo
  {journal} {Eur. Phys. J. C}\ }\textbf {\bibinfo {volume} {78}},\ \bibinfo
  {pages} {473} (\bibinfo {year} {2018})},\ \Eprint
  {http://arxiv.org/abs/1804.01019} {arXiv:1804.01019 [hep-ex]} \BibitemShut
  {NoStop}%
\bibitem [{\citenamefont {Schwanda}\ \emph {et~al.}(2008)\citenamefont
  {Schwanda} \emph {et~al.}}]{Schwanda:2008kw}%
  \BibitemOpen
  \bibfield  {author} {\bibinfo {author} {\bibfnamefont {C.}~\bibnamefont
  {Schwanda}} \emph {et~al.} (\bibinfo {collaboration} {Belle}),\ }\href
  {\doibase 10.1103/PhysRevD.78.032016} {\bibfield  {journal} {\bibinfo
  {journal} {Phys. Rev. D}\ }\textbf {\bibinfo {volume} {78}},\ \bibinfo
  {pages} {032016} (\bibinfo {year} {2008})},\ \Eprint
  {http://arxiv.org/abs/0803.2158} {arXiv:0803.2158 [hep-ex]} \BibitemShut
  {NoStop}%
\bibitem [{\citenamefont {Aubert}\ \emph {et~al.}(2010)\citenamefont {Aubert}
  \emph {et~al.}}]{Aubert:2009qda}%
  \BibitemOpen
  \bibfield  {author} {\bibinfo {author} {\bibfnamefont {B.}~\bibnamefont
  {Aubert}} \emph {et~al.} (\bibinfo {collaboration} {BaBar}),\ }\href
  {\doibase 10.1103/PhysRevD.81.032003} {\bibfield  {journal} {\bibinfo
  {journal} {Phys. Rev. D}\ }\textbf {\bibinfo {volume} {81}},\ \bibinfo
  {pages} {032003} (\bibinfo {year} {2010})},\ \Eprint
  {http://arxiv.org/abs/0908.0415} {arXiv:0908.0415 [hep-ex]} \BibitemShut
  {NoStop}%
\bibitem [{\citenamefont {Bilenky}\ \emph {et~al.}(1995)\citenamefont
  {Bilenky}, \citenamefont {Rodrigo},\ and\ \citenamefont
  {Santamaria}}]{Bilenky:1994ad}%
  \BibitemOpen
  \bibfield  {author} {\bibinfo {author} {\bibfnamefont {M.~S.}\ \bibnamefont
  {Bilenky}}, \bibinfo {author} {\bibfnamefont {G.}~\bibnamefont {Rodrigo}}, \
  and\ \bibinfo {author} {\bibfnamefont {A.}~\bibnamefont {Santamaria}},\
  }\href {\doibase 10.1016/0550-3213(94)00586-4} {\bibfield  {journal}
  {\bibinfo  {journal} {Nucl. Phys. B}\ }\textbf {\bibinfo {volume} {439}},\
  \bibinfo {pages} {505} (\bibinfo {year} {1995})},\ \Eprint
  {http://arxiv.org/abs/hep-ph/9410258} {arXiv:hep-ph/9410258} \BibitemShut
  {NoStop}%
\bibitem [{\citenamefont {Rodrigo}\ \emph {et~al.}(1997)\citenamefont
  {Rodrigo}, \citenamefont {Santamaria},\ and\ \citenamefont
  {Bilenky}}]{Rodrigo:1997gy}%
  \BibitemOpen
  \bibfield  {author} {\bibinfo {author} {\bibfnamefont {G.}~\bibnamefont
  {Rodrigo}}, \bibinfo {author} {\bibfnamefont {A.}~\bibnamefont {Santamaria}},
  \ and\ \bibinfo {author} {\bibfnamefont {M.~S.}\ \bibnamefont {Bilenky}},\
  }\href {\doibase 10.1103/PhysRevLett.79.193} {\bibfield  {journal} {\bibinfo
  {journal} {Phys. Rev. Lett.}\ }\textbf {\bibinfo {volume} {79}},\ \bibinfo
  {pages} {193} (\bibinfo {year} {1997})},\ \Eprint
  {http://arxiv.org/abs/hep-ph/9703358} {arXiv:hep-ph/9703358 [hep-ph]}
  \BibitemShut {NoStop}%
%%CITATION = HEP-PH/9703358;%%
\bibitem [{\citenamefont {Bilenky}\ \emph {et~al.}(1999)\citenamefont
  {Bilenky}, \citenamefont {Caberera}, \citenamefont {Fuster}, \citenamefont
  {Marti}, \citenamefont {Rodrigo},\ and\ \citenamefont
  {Santamaria}}]{Bilenky:1998nk}%
  \BibitemOpen
  \bibfield  {author} {\bibinfo {author} {\bibfnamefont {M.~S.}\ \bibnamefont
  {Bilenky}}, \bibinfo {author} {\bibfnamefont {S.}~\bibnamefont {Caberera}},
  \bibinfo {author} {\bibfnamefont {J.}~\bibnamefont {Fuster}}, \bibinfo
  {author} {\bibfnamefont {S.}~\bibnamefont {Marti}}, \bibinfo {author}
  {\bibfnamefont {G.}~\bibnamefont {Rodrigo}}, \ and\ \bibinfo {author}
  {\bibfnamefont {A.}~\bibnamefont {Santamaria}},\ }\href {\doibase
  10.1103/PhysRevD.60.114006} {\bibfield  {journal} {\bibinfo  {journal} {Phys.
  Rev. D}\ }\textbf {\bibinfo {volume} {60}},\ \bibinfo {pages} {114006}
  (\bibinfo {year} {1999})},\ \Eprint {http://arxiv.org/abs/hep-ph/9807489}
  {arXiv:hep-ph/9807489} \BibitemShut {NoStop}%
\bibitem [{\citenamefont {Rodrigo}\ \emph {et~al.}(1999)\citenamefont
  {Rodrigo}, \citenamefont {Bilenky},\ and\ \citenamefont
  {Santamaria}}]{Rodrigo:1999qg}%
  \BibitemOpen
  \bibfield  {author} {\bibinfo {author} {\bibfnamefont {G.}~\bibnamefont
  {Rodrigo}}, \bibinfo {author} {\bibfnamefont {M.~S.}\ \bibnamefont
  {Bilenky}}, \ and\ \bibinfo {author} {\bibfnamefont {A.}~\bibnamefont
  {Santamaria}},\ }\href {\doibase 10.1016/S0550-3213(99)00293-X} {\bibfield
  {journal} {\bibinfo  {journal} {Nucl. Phys. B}\ }\textbf {\bibinfo {volume}
  {554}},\ \bibinfo {pages} {257} (\bibinfo {year} {1999})},\ \Eprint
  {http://arxiv.org/abs/hep-ph/9905276} {arXiv:hep-ph/9905276} \BibitemShut
  {NoStop}%
\bibitem [{\citenamefont {Bernreuther}\ \emph {et~al.}(1997)\citenamefont
  {Bernreuther}, \citenamefont {Brandenburg},\ and\ \citenamefont
  {Uwer}}]{Bernreuther:1997jn}%
  \BibitemOpen
  \bibfield  {author} {\bibinfo {author} {\bibfnamefont {W.}~\bibnamefont
  {Bernreuther}}, \bibinfo {author} {\bibfnamefont {A.}~\bibnamefont
  {Brandenburg}}, \ and\ \bibinfo {author} {\bibfnamefont {P.}~\bibnamefont
  {Uwer}},\ }\href {\doibase 10.1103/PhysRevLett.79.189} {\bibfield  {journal}
  {\bibinfo  {journal} {Phys. Rev. Lett.}\ }\textbf {\bibinfo {volume} {79}},\
  \bibinfo {pages} {189} (\bibinfo {year} {1997})},\ \Eprint
  {http://arxiv.org/abs/hep-ph/9703305} {arXiv:hep-ph/9703305 [hep-ph]}
  \BibitemShut {NoStop}%
%%CITATION = HEP-PH/9703305;%%
\bibitem [{\citenamefont {Brandenburg}\ and\ \citenamefont
  {Uwer}(1998)}]{Brandenburg:1997pu}%
  \BibitemOpen
  \bibfield  {author} {\bibinfo {author} {\bibfnamefont {A.}~\bibnamefont
  {Brandenburg}}\ and\ \bibinfo {author} {\bibfnamefont {P.}~\bibnamefont
  {Uwer}},\ }\href {\doibase 10.1016/S0550-3213(97)00790-6} {\bibfield
  {journal} {\bibinfo  {journal} {Nucl. Phys. B}\ }\textbf {\bibinfo {volume}
  {515}},\ \bibinfo {pages} {279} (\bibinfo {year} {1998})},\ \Eprint
  {http://arxiv.org/abs/hep-ph/9708350} {arXiv:hep-ph/9708350} \BibitemShut
  {NoStop}%
\bibitem [{\citenamefont {Nason}\ and\ \citenamefont
  {Oleari}(1997)}]{Nason:1997tz}%
  \BibitemOpen
  \bibfield  {author} {\bibinfo {author} {\bibfnamefont {P.}~\bibnamefont
  {Nason}}\ and\ \bibinfo {author} {\bibfnamefont {C.}~\bibnamefont {Oleari}},\
  }\href {\doibase 10.1016/S0370-2693(97)00721-1} {\bibfield  {journal}
  {\bibinfo  {journal} {Phys. Lett. B}\ }\textbf {\bibinfo {volume} {407}},\
  \bibinfo {pages} {57} (\bibinfo {year} {1997})},\ \Eprint
  {http://arxiv.org/abs/hep-ph/9705295} {arXiv:hep-ph/9705295} \BibitemShut
  {NoStop}%
\bibitem [{\citenamefont {Nason}\ and\ \citenamefont
  {Oleari}(1998)}]{Nason:1997nw}%
  \BibitemOpen
  \bibfield  {author} {\bibinfo {author} {\bibfnamefont {P.}~\bibnamefont
  {Nason}}\ and\ \bibinfo {author} {\bibfnamefont {C.}~\bibnamefont {Oleari}},\
  }\href {\doibase 10.1016/S0550-3213(98)00125-4} {\bibfield  {journal}
  {\bibinfo  {journal} {Nucl. Phys. B}\ }\textbf {\bibinfo {volume} {521}},\
  \bibinfo {pages} {237} (\bibinfo {year} {1998})},\ \Eprint
  {http://arxiv.org/abs/hep-ph/9709360} {arXiv:hep-ph/9709360} \BibitemShut
  {NoStop}%
\bibitem [{\citenamefont {Gehrmann-De~Ridder}\ \emph
  {et~al.}(2008)\citenamefont {Gehrmann-De~Ridder}, \citenamefont {Gehrmann},
  \citenamefont {Glover},\ and\ \citenamefont
  {Heinrich}}]{GehrmannDeRidder:2008ug}%
  \BibitemOpen
  \bibfield  {author} {\bibinfo {author} {\bibfnamefont {A.}~\bibnamefont
  {Gehrmann-De~Ridder}}, \bibinfo {author} {\bibfnamefont {T.}~\bibnamefont
  {Gehrmann}}, \bibinfo {author} {\bibfnamefont {E.~W.~N.}\ \bibnamefont
  {Glover}}, \ and\ \bibinfo {author} {\bibfnamefont {G.}~\bibnamefont
  {Heinrich}},\ }\href {\doibase 10.1103/PhysRevLett.100.172001} {\bibfield
  {journal} {\bibinfo  {journal} {Phys. Rev. Lett.}\ }\textbf {\bibinfo
  {volume} {100}},\ \bibinfo {pages} {172001} (\bibinfo {year} {2008})},\
  \Eprint {http://arxiv.org/abs/0802.0813} {arXiv:0802.0813 [hep-ph]}
  \BibitemShut {NoStop}%
\bibitem [{\citenamefont {Abreu}\ \emph {et~al.}(1998)\citenamefont {Abreu}
  \emph {et~al.}}]{Abreu:1997ey}%
  \BibitemOpen
  \bibfield  {author} {\bibinfo {author} {\bibfnamefont {P.}~\bibnamefont
  {Abreu}} \emph {et~al.} (\bibinfo {collaboration} {DELPHI}),\ }\href
  {\doibase 10.1016/S0370-2693(97)01442-1} {\bibfield  {journal} {\bibinfo
  {journal} {Phys. Lett.}\ }\textbf {\bibinfo {volume} {B418}},\ \bibinfo
  {pages} {430} (\bibinfo {year} {1998})}\BibitemShut {NoStop}%
%%CITATION = PHLTA,B418,430;%%
\bibitem [{\citenamefont {Brandenburg}\ \emph {et~al.}(1999)\citenamefont
  {Brandenburg}, \citenamefont {Burrows}, \citenamefont {Muller}, \citenamefont
  {Oishi},\ and\ \citenamefont {Uwer}}]{Brandenburg:1999nb}%
  \BibitemOpen
  \bibfield  {author} {\bibinfo {author} {\bibfnamefont {A.}~\bibnamefont
  {Brandenburg}}, \bibinfo {author} {\bibfnamefont {P.~N.}\ \bibnamefont
  {Burrows}}, \bibinfo {author} {\bibfnamefont {D.}~\bibnamefont {Muller}},
  \bibinfo {author} {\bibfnamefont {N.}~\bibnamefont {Oishi}}, \ and\ \bibinfo
  {author} {\bibfnamefont {P.}~\bibnamefont {Uwer}},\ }\href {\doibase
  10.1016/S0370-2693(99)01194-6} {\bibfield  {journal} {\bibinfo  {journal}
  {Phys. Lett.}\ }\textbf {\bibinfo {volume} {B468}},\ \bibinfo {pages} {168}
  (\bibinfo {year} {1999})},\ \Eprint {http://arxiv.org/abs/hep-ph/9905495}
  {arXiv:hep-ph/9905495 [hep-ph]} \BibitemShut {NoStop}%
%%CITATION = HEP-PH/9905495;%%
\bibitem [{\citenamefont {Abe}\ \emph {et~al.}(1999)\citenamefont {Abe} \emph
  {et~al.}}]{Abe:1998kr}%
  \BibitemOpen
  \bibfield  {author} {\bibinfo {author} {\bibfnamefont {K.}~\bibnamefont
  {Abe}} \emph {et~al.} (\bibinfo {collaboration} {SLD}),\ }\href {\doibase
  10.1103/PhysRevD.59.012002} {\bibfield  {journal} {\bibinfo  {journal} {Phys.
  Rev.}\ }\textbf {\bibinfo {volume} {D59}},\ \bibinfo {pages} {012002}
  (\bibinfo {year} {1999})},\ \Eprint {http://arxiv.org/abs/hep-ex/9805023}
  {arXiv:hep-ex/9805023 [hep-ex]} \BibitemShut {NoStop}%
%%CITATION = HEP-EX/9805023;%%
\bibitem [{\citenamefont {Barate}\ \emph {et~al.}(2000)\citenamefont {Barate}
  \emph {et~al.}}]{Barate:2000ab}%
  \BibitemOpen
  \bibfield  {author} {\bibinfo {author} {\bibfnamefont {R.}~\bibnamefont
  {Barate}} \emph {et~al.} (\bibinfo {collaboration} {ALEPH}),\ }\href
  {\doibase 10.1007/s100520000533} {\bibfield  {journal} {\bibinfo  {journal}
  {Eur. Phys. J.}\ }\textbf {\bibinfo {volume} {C18}},\ \bibinfo {pages} {1}
  (\bibinfo {year} {2000})},\ \Eprint {http://arxiv.org/abs/hep-ex/0008013}
  {arXiv:hep-ex/0008013 [hep-ex]} \BibitemShut {NoStop}%
%%CITATION = HEP-EX/0008013;%%
\bibitem [{\citenamefont {Abbiendi}\ \emph {et~al.}(2001)\citenamefont
  {Abbiendi} \emph {et~al.}}]{Abbiendi:2001tw}%
  \BibitemOpen
  \bibfield  {author} {\bibinfo {author} {\bibfnamefont {G.}~\bibnamefont
  {Abbiendi}} \emph {et~al.} (\bibinfo {collaboration} {OPAL}),\ }\href
  {\doibase 10.1007/100520100746} {\bibfield  {journal} {\bibinfo  {journal}
  {Eur. Phys. J.}\ }\textbf {\bibinfo {volume} {C21}},\ \bibinfo {pages} {411}
  (\bibinfo {year} {2001})},\ \Eprint {http://arxiv.org/abs/hep-ex/0105046}
  {arXiv:hep-ex/0105046 [hep-ex]} \BibitemShut {NoStop}%
%%CITATION = HEP-EX/0105046;%%
\bibitem [{\citenamefont {Abdallah}\ \emph {et~al.}(2006)\citenamefont
  {Abdallah} \emph {et~al.}}]{Abdallah:2005cv}%
  \BibitemOpen
  \bibfield  {author} {\bibinfo {author} {\bibfnamefont {J.}~\bibnamefont
  {Abdallah}} \emph {et~al.} (\bibinfo {collaboration} {DELPHI}),\ }\href
  {\doibase 10.1140/epjc/s2006-02497-6} {\bibfield  {journal} {\bibinfo
  {journal} {Eur. Phys. J.}\ }\textbf {\bibinfo {volume} {C46}},\ \bibinfo
  {pages} {569} (\bibinfo {year} {2006})},\ \Eprint
  {http://arxiv.org/abs/hep-ex/0603046} {arXiv:hep-ex/0603046 [hep-ex]}
  \BibitemShut {NoStop}%
%%CITATION = HEP-EX/0603046;%%
\bibitem [{\citenamefont {Abdallah}\ \emph {et~al.}(2008)\citenamefont
  {Abdallah} \emph {et~al.}}]{Abdallah:2008ac}%
  \BibitemOpen
  \bibfield  {author} {\bibinfo {author} {\bibfnamefont {J.}~\bibnamefont
  {Abdallah}} \emph {et~al.} (\bibinfo {collaboration} {DELPHI}),\ }\href
  {\doibase 10.1140/epjc/s10052-008-0631-5} {\bibfield  {journal} {\bibinfo
  {journal} {Eur. Phys. J.}\ }\textbf {\bibinfo {volume} {C55}},\ \bibinfo
  {pages} {525} (\bibinfo {year} {2008})},\ \Eprint
  {http://arxiv.org/abs/0804.3883} {arXiv:0804.3883 [hep-ex]} \BibitemShut
  {NoStop}%
%%CITATION = ARXIV:0804.3883;%%
\bibitem [{\citenamefont {Kluth}(2006)}]{Kluth:2006bw}%
  \BibitemOpen
  \bibfield  {author} {\bibinfo {author} {\bibfnamefont {S.}~\bibnamefont
  {Kluth}},\ }\href {\doibase 10.1088/0034-4885/69/6/R04} {\bibfield  {journal}
  {\bibinfo  {journal} {Rept. Prog. Phys.}\ }\textbf {\bibinfo {volume} {69}},\
  \bibinfo {pages} {1771} (\bibinfo {year} {2006})},\ \Eprint
  {http://arxiv.org/abs/hep-ex/0603011} {arXiv:hep-ex/0603011} \BibitemShut
  {NoStop}%
\bibitem [{\citenamefont {Aparisi}\ \emph {et~al.}(2022)\citenamefont {Aparisi}
  \emph {et~al.}}]{Aparisi:2021tym}%
  \BibitemOpen
  \bibfield  {author} {\bibinfo {author} {\bibfnamefont {J.}~\bibnamefont
  {Aparisi}} \emph {et~al.},\ }\href {\doibase 10.1103/PhysRevLett.128.122001}
  {\bibfield  {journal} {\bibinfo  {journal} {Phys. Rev. Lett.}\ }\textbf
  {\bibinfo {volume} {128}},\ \bibinfo {pages} {122001} (\bibinfo {year}
  {2022})},\ \Eprint {http://arxiv.org/abs/2110.10202} {arXiv:2110.10202
  [hep-ph]} \BibitemShut {NoStop}%
\bibitem [{\citenamefont {{ATLAS Collaboration}}(2020)}]{ATLAS-CONF-2020-027}%
  \BibitemOpen
  \bibfield  {author} {\bibinfo {author} {\bibnamefont {{ATLAS
  Collaboration}}},\ }\href {https://cds.cern.ch/record/2725733} {\bibfield
  {journal} {\bibinfo  {journal} {ATLAS-CONF-2020-027}\ } (\bibinfo {year}
  {2020})}\BibitemShut {NoStop}%
\bibitem [{\citenamefont {{CMS collaboration}}(2019)}]{Sirunyan:2018koj}%
  \BibitemOpen
  \bibfield  {author} {\bibinfo {author} {\bibnamefont {{CMS collaboration}}},\
  }\href {\doibase 10.1140/epjc/s10052-019-6909-y} {\bibfield  {journal}
  {\bibinfo  {journal} {Eur. Phys. J.}\ }\textbf {\bibinfo {volume} {C79}},\
  \bibinfo {pages} {421} (\bibinfo {year} {2019})},\ \Eprint
  {http://arxiv.org/abs/1809.10733} {arXiv:1809.10733 [hep-ex]} \BibitemShut
  {NoStop}%
%%CITATION = ARXIV:1809.10733;%%
\bibitem [{\citenamefont {Mateu}\ and\ \citenamefont
  {Ortega}(2018)}]{Mateu:2017hlz}%
  \BibitemOpen
  \bibfield  {author} {\bibinfo {author} {\bibfnamefont {V.}~\bibnamefont
  {Mateu}}\ and\ \bibinfo {author} {\bibfnamefont {P.~G.}\ \bibnamefont
  {Ortega}},\ }\href {\doibase 10.1007/JHEP01(2018)122} {\bibfield  {journal}
  {\bibinfo  {journal} {JHEP}\ } {\bibinfo {volume} {01}},\ \bibinfo
  {pages} {122} (\bibinfo {year} {2018})},\ \Eprint
  {http://arxiv.org/abs/1711.05755} {arXiv:1711.05755 [hep-ph]} \BibitemShut
  {NoStop}%
\bibitem [{\citenamefont {Dehnadi}\ \emph {et~al.}(2013)\citenamefont
  {Dehnadi}, \citenamefont {Hoang}, \citenamefont {Mateu},\ and\ \citenamefont
  {Zebarjad}}]{Dehnadi:2011gc}%
  \BibitemOpen
  \bibfield  {author} {\bibinfo {author} {\bibfnamefont {B.}~\bibnamefont
  {Dehnadi}}, \bibinfo {author} {\bibfnamefont {A.~H.}\ \bibnamefont {Hoang}},
  \bibinfo {author} {\bibfnamefont {V.}~\bibnamefont {Mateu}}, \ and\ \bibinfo
  {author} {\bibfnamefont {S.~M.}\ \bibnamefont {Zebarjad}},\ }\href {\doibase
  10.1007/JHEP09(2013)103} {\bibfield  {journal} {\bibinfo  {journal} {JHEP}\
  }\textbf {\bibinfo {volume} {09}},\ \bibinfo {pages} {103} (\bibinfo {year}
  {2013})},\ \Eprint {http://arxiv.org/abs/1102.2264} {arXiv:1102.2264
  [hep-ph]} \BibitemShut {NoStop}%
\bibitem [{\citenamefont {Aubert}\ \emph {et~al.}(2009)\citenamefont {Aubert}
  \emph {et~al.}}]{BaBar:2008cmq}%
  \BibitemOpen
  \bibfield  {author} {\bibinfo {author} {\bibfnamefont {B.}~\bibnamefont
  {Aubert}} \emph {et~al.} (\bibinfo {collaboration} {BaBar}),\ }\href
  {\doibase 10.1103/PhysRevLett.102.012001} {\bibfield  {journal} {\bibinfo
  {journal} {Phys. Rev. Lett.}\ }\textbf {\bibinfo {volume} {102}},\ \bibinfo
  {pages} {012001} (\bibinfo {year} {2009})},\ \Eprint
  {http://arxiv.org/abs/0809.4120} {arXiv:0809.4120 [hep-ex]} \BibitemShut
  {NoStop}%
\bibitem [{\citenamefont {Hoang}\ \emph {et~al.}(2012)\citenamefont {Hoang},
  \citenamefont {Ruiz-Femenia},\ and\ \citenamefont
  {Stahlhofen}}]{Hoang:2012us}%
  \BibitemOpen
  \bibfield  {author} {\bibinfo {author} {\bibfnamefont {A.}~\bibnamefont
  {Hoang}}, \bibinfo {author} {\bibfnamefont {P.}~\bibnamefont {Ruiz-Femenia}},
  \ and\ \bibinfo {author} {\bibfnamefont {M.}~\bibnamefont {Stahlhofen}},\
  }\href {\doibase 10.1007/JHEP10(2012)188} {\bibfield  {journal} {\bibinfo
  {journal} {JHEP}\ } {\bibinfo {volume} {10}},\ \bibinfo {pages} {188}
  (\bibinfo {year} {2012})},\ \Eprint {http://arxiv.org/abs/1209.0450}
  {arXiv:1209.0450 [hep-ph]} \BibitemShut {NoStop}%
%%CITATION = ARXIV:1209.0450;%%
\bibitem [{\citenamefont {Boito}\ \emph
  {et~al.}(2021{\natexlab{a}})\citenamefont {Boito}, \citenamefont {Mateu},\
  and\ \citenamefont {Rodrigues}}]{Boito:2021wbj}%
  \BibitemOpen
  \bibfield  {author} {\bibinfo {author} {\bibfnamefont {D.}~\bibnamefont
  {Boito}}, \bibinfo {author} {\bibfnamefont {V.}~\bibnamefont {Mateu}}, \ and\
  \bibinfo {author} {\bibfnamefont {M.~V.}\ \bibnamefont {Rodrigues}},\ }\href
  {\doibase 10.1007/JHEP08(2021)027} {\bibfield  {journal} {\bibinfo  {journal}
  {JHEP}\ }\textbf {\bibinfo {volume} {08}},\ \bibinfo {pages} {027} (\bibinfo
  {year} {2021}{\natexlab{a}})},\ \Eprint {http://arxiv.org/abs/2106.05660}
  {arXiv:2106.05660 [hep-ph]} \BibitemShut {NoStop}%
\bibitem [{\citenamefont {Fuster}\ \emph {et~al.}()\citenamefont {Fuster},
  \citenamefont {Irles}, \citenamefont {Tairafune}, \citenamefont {Rodrigo},
  \citenamefont {Vos}, \citenamefont {Yamamoto},\ and\ \citenamefont
  {Yonamine}}]{ILDnote2020}%
  \BibitemOpen
  \bibfield  {author} {\bibinfo {author} {\bibfnamefont {J.}~\bibnamefont
  {Fuster}}, \bibinfo {author} {\bibfnamefont {A.}~\bibnamefont {Irles}},
  \bibinfo {author} {\bibfnamefont {S.}~\bibnamefont {Tairafune}}, \bibinfo
  {author} {\bibfnamefont {G.}~\bibnamefont {Rodrigo}}, \bibinfo {author}
  {\bibfnamefont {M.}~\bibnamefont {Vos}}, \bibinfo {author} {\bibfnamefont
  {H.}~\bibnamefont {Yamamoto}}, \ and\ \bibinfo {author} {\bibfnamefont
  {R.}~\bibnamefont {Yonamine}},\ }\href@noop {} {\bibfield  {journal}
  {\bibinfo  {journal} {ILD note}\ }}\Eprint
  {http://arxiv.org/abs/ILD-PHYS-PUB--2021-001} {ILD-PHYS-PUB--2021-001}
  \BibitemShut {NoStop}%
\bibitem [{\citenamefont {Bernreuther}\ \emph {et~al.}(2018)\citenamefont
  {Bernreuther}, \citenamefont {Chen},\ and\ \citenamefont
  {Si}}]{Bernreuther:2018ynm}%
  \BibitemOpen
  \bibfield  {author} {\bibinfo {author} {\bibfnamefont {W.}~\bibnamefont
  {Bernreuther}}, \bibinfo {author} {\bibfnamefont {L.}~\bibnamefont {Chen}}, \
  and\ \bibinfo {author} {\bibfnamefont {Z.-G.}\ \bibnamefont {Si}},\ }\href
  {\doibase 10.1007/JHEP07(2018)159} {\bibfield  {journal} {\bibinfo  {journal}
  {JHEP}\ }\textbf {\bibinfo {volume} {07}},\ \bibinfo {pages} {159} (\bibinfo
  {year} {2018})},\ \Eprint {http://arxiv.org/abs/1805.06658} {arXiv:1805.06658
  [hep-ph]} \BibitemShut {NoStop}%
\bibitem [{\citenamefont {Schael}\ \emph {et~al.}(2006)\citenamefont {Schael}
  \emph {et~al.}}]{ALEPH:2005ab}%
  \BibitemOpen
  \bibfield  {author} {\bibinfo {author} {\bibfnamefont {S.}~\bibnamefont
  {Schael}} \emph {et~al.} (\bibinfo {collaboration} {ALEPH, DELPHI, L3, OPAL,
  SLD, LEP Electroweak Working Group, SLD Electroweak Group, SLD Heavy Flavour
  Group}),\ }\href {\doibase 10.1016/j.physrep.2005.12.006} {\bibfield
  {journal} {\bibinfo  {journal} {Phys. Rept.}\ }\textbf {\bibinfo {volume}
  {427}},\ \bibinfo {pages} {257} (\bibinfo {year} {2006})},\ \Eprint
  {http://arxiv.org/abs/hep-ex/0509008} {arXiv:hep-ex/0509008 [hep-ex]}
  \BibitemShut {NoStop}%
%%CITATION = HEP-EX/0509008;%%
\bibitem [{\citenamefont {Kluth}(2022)}]{Kluth:2022ucw}%
  \BibitemOpen
  \bibfield  {author} {\bibinfo {author} {\bibfnamefont {S.}~\bibnamefont
  {Kluth}},\ }\href@noop {} {\  (\bibinfo {year} {2022})},\ \Eprint
  {http://arxiv.org/abs/2201.02417} {arXiv:2201.02417 [hep-ph]} \BibitemShut
  {NoStop}%
\bibitem [{\citenamefont {Abada}\ \emph
  {et~al.}(2019{\natexlab{a}})\citenamefont {Abada} \emph
  {et~al.}}]{FCC:2018byv}%
  \BibitemOpen
  \bibfield  {author} {\bibinfo {author} {\bibfnamefont {A.}~\bibnamefont
  {Abada}} \emph {et~al.} (\bibinfo {collaboration} {FCC}),\ }\href {\doibase
  10.1140/epjc/s10052-019-6904-3} {\bibfield  {journal} {\bibinfo  {journal}
  {Eur. Phys. J. C}\ }\textbf {\bibinfo {volume} {79}},\ \bibinfo {pages} {474}
  (\bibinfo {year} {2019}{\natexlab{a}})}\BibitemShut {NoStop}%
\bibitem [{\citenamefont {Abada}\ \emph
  {et~al.}(2019{\natexlab{b}})\citenamefont {Abada} \emph
  {et~al.}}]{FCC:2018evy}%
  \BibitemOpen
  \bibfield  {author} {\bibinfo {author} {\bibfnamefont {A.}~\bibnamefont
  {Abada}} \emph {et~al.} (\bibinfo {collaboration} {FCC}),\ }\href {\doibase
  10.1140/epjst/e2019-900045-4} {\bibfield  {journal} {\bibinfo  {journal}
  {Eur. Phys. J. ST}\ }\textbf {\bibinfo {volume} {228}},\ \bibinfo {pages}
  {261} (\bibinfo {year} {2019}{\natexlab{b}})}\BibitemShut {NoStop}%
\bibitem [{\citenamefont {de~Blas}\ \emph {et~al.}(2020)\citenamefont {de~Blas}
  \emph {et~al.}}]{deBlas:2019rxi}%
  \BibitemOpen
  \bibfield  {author} {\bibinfo {author} {\bibfnamefont {J.}~\bibnamefont
  {de~Blas}} \emph {et~al.},\ }\href {\doibase 10.1007/JHEP01(2020)139}
  {\bibfield  {journal} {\bibinfo  {journal} {JHEP}\ }\textbf {\bibinfo
  {volume} {01}},\ \bibinfo {pages} {139} (\bibinfo {year} {2020})},\ \Eprint
  {http://arxiv.org/abs/1905.03764} {arXiv:1905.03764 [hep-ph]} \BibitemShut
  {NoStop}%
\bibitem [{\citenamefont {Cepeda}\ \emph {et~al.}(2019)\citenamefont {Cepeda}
  \emph {et~al.}}]{Cepeda:2019klc}%
  \BibitemOpen
  \bibfield  {author} {\bibinfo {author} {\bibfnamefont {M.}~\bibnamefont
  {Cepeda}} \emph {et~al.},\ }\href {\doibase 10.23731/CYRM-2019-007.221}
  {\bibfield  {journal} {\bibinfo  {journal} {CERN Yellow Rep. Monogr.}\
  }\textbf {\bibinfo {volume} {7}},\ \bibinfo {pages} {221} (\bibinfo {year}
  {2019})},\ \Eprint {http://arxiv.org/abs/1902.00134} {arXiv:1902.00134
  [hep-ph]} \BibitemShut {NoStop}%
%%CITATION = ARXIV:1902.00134;%%
\bibitem [{\citenamefont {Yan}\ \emph {et~al.}(2016)\citenamefont {Yan},
  \citenamefont {Watanuki}, \citenamefont {Fujii}, \citenamefont {Ishikawa},
  \citenamefont {Jeans}, \citenamefont {Strube}, \citenamefont {Tian},\ and\
  \citenamefont {Yamamoto}}]{Yan:2016xyx}%
  \BibitemOpen
  \bibfield  {author} {\bibinfo {author} {\bibfnamefont {J.}~\bibnamefont
  {Yan}}, \bibinfo {author} {\bibfnamefont {S.}~\bibnamefont {Watanuki}},
  \bibinfo {author} {\bibfnamefont {K.}~\bibnamefont {Fujii}}, \bibinfo
  {author} {\bibfnamefont {A.}~\bibnamefont {Ishikawa}}, \bibinfo {author}
  {\bibfnamefont {D.}~\bibnamefont {Jeans}}, \bibinfo {author} {\bibfnamefont
  {J.}~\bibnamefont {Strube}}, \bibinfo {author} {\bibfnamefont
  {J.}~\bibnamefont {Tian}}, \ and\ \bibinfo {author} {\bibfnamefont
  {H.}~\bibnamefont {Yamamoto}},\ }\href {\doibase 10.1103/PhysRevD.94.113002}
  {\bibfield  {journal} {\bibinfo  {journal} {Phys. Rev. D}\ }\textbf {\bibinfo
  {volume} {94}},\ \bibinfo {pages} {113002} (\bibinfo {year} {2016})},\
  \bibinfo {note} {[Erratum: Phys.Rev.D 103, 099903 (2021)]},\ \Eprint
  {http://arxiv.org/abs/1604.07524} {arXiv:1604.07524 [hep-ex]} \BibitemShut
  {NoStop}%
\bibitem [{\citenamefont {de~Florian}\ \emph {et~al.}(2016)\citenamefont
  {de~Florian} \emph {et~al.}}]{deFlorian:2016spz}%
  \BibitemOpen
  \bibfield  {author} {\bibinfo {author} {\bibfnamefont {D.}~\bibnamefont
  {de~Florian}} \emph {et~al.} (\bibinfo {collaboration} {LHC Higgs Cross
  Section Working Group}),\ }\href {\doibase 10.23731/CYRM-2017-002} {\ \textbf
  {\bibinfo {volume} {2/2017}} (\bibinfo {year} {2016}),\
  10.23731/CYRM-2017-002},\ \Eprint {http://arxiv.org/abs/1610.07922}
  {arXiv:1610.07922 [hep-ph]} \BibitemShut {NoStop}%
\bibitem [{\citenamefont {Barklow}\ \emph {et~al.}(2018)\citenamefont
  {Barklow}, \citenamefont {Fujii}, \citenamefont {Jung}, \citenamefont {Karl},
  \citenamefont {List}, \citenamefont {Ogawa}, \citenamefont {Peskin},\ and\
  \citenamefont {Tian}}]{Barklow:2017suo}%
  \BibitemOpen
  \bibfield  {author} {\bibinfo {author} {\bibfnamefont {T.}~\bibnamefont
  {Barklow}}, \bibinfo {author} {\bibfnamefont {K.}~\bibnamefont {Fujii}},
  \bibinfo {author} {\bibfnamefont {S.}~\bibnamefont {Jung}}, \bibinfo {author}
  {\bibfnamefont {R.}~\bibnamefont {Karl}}, \bibinfo {author} {\bibfnamefont
  {J.}~\bibnamefont {List}}, \bibinfo {author} {\bibfnamefont {T.}~\bibnamefont
  {Ogawa}}, \bibinfo {author} {\bibfnamefont {M.~E.}\ \bibnamefont {Peskin}}, \
  and\ \bibinfo {author} {\bibfnamefont {J.}~\bibnamefont {Tian}},\ }\href
  {\doibase 10.1103/PhysRevD.97.053003} {\bibfield  {journal} {\bibinfo
  {journal} {Phys. Rev. D}\ }\textbf {\bibinfo {volume} {97}},\ \bibinfo
  {pages} {053003} (\bibinfo {year} {2018})},\ \Eprint
  {http://arxiv.org/abs/1708.08912} {arXiv:1708.08912 [hep-ph]} \BibitemShut
  {NoStop}%
\bibitem [{\citenamefont {Abramowicz}\ \emph {et~al.}(2017)\citenamefont
  {Abramowicz} \emph {et~al.}}]{Abramowicz:2016zbo}%
  \BibitemOpen
  \bibfield  {author} {\bibinfo {author} {\bibfnamefont {H.}~\bibnamefont
  {Abramowicz}} \emph {et~al.},\ }\href {\doibase
  10.1140/epjc/s10052-017-4968-5} {\bibfield  {journal} {\bibinfo  {journal}
  {Eur. Phys. J. C}\ }\textbf {\bibinfo {volume} {77}},\ \bibinfo {pages} {475}
  (\bibinfo {year} {2017})},\ \Eprint {http://arxiv.org/abs/1608.07538}
  {arXiv:1608.07538 [hep-ex]} \BibitemShut {NoStop}%
\bibitem [{\citenamefont {Fujii}\ \emph {et~al.}(2019)\citenamefont {Fujii}
  \emph {et~al.}}]{Fujii:2019zll}%
  \BibitemOpen
  \bibfield  {author} {\bibinfo {author} {\bibfnamefont {K.}~\bibnamefont
  {Fujii}} \emph {et~al.} (\bibinfo {collaboration} {LCC Physics Working
  Group}),\ }\href@noop {} {\  (\bibinfo {year} {2019})},\ \Eprint
  {http://arxiv.org/abs/1908.11299} {arXiv:1908.11299 [hep-ex]} \BibitemShut
  {NoStop}%
\bibitem [{\citenamefont {Bambade}\ \emph {et~al.}(2019)\citenamefont {Bambade}
  \emph {et~al.}}]{Bambade:2019fyw}%
  \BibitemOpen
  \bibfield  {author} {\bibinfo {author} {\bibfnamefont {P.}~\bibnamefont
  {Bambade}} \emph {et~al.},\ }\href@noop {} {\  (\bibinfo {year} {2019})},\
  \Eprint {http://arxiv.org/abs/1903.01629} {arXiv:1903.01629 [hep-ex]}
  \BibitemShut {NoStop}%
\bibitem [{\citenamefont {Aoki}\ \emph {et~al.}(2021)\citenamefont {Aoki} \emph
  {et~al.}}]{Aoki:2021kgd}%
  \BibitemOpen
  \bibfield  {author} {\bibinfo {author} {\bibfnamefont {Y.}~\bibnamefont
  {Aoki}} \emph {et~al.},\ }\href@noop {} {\  (\bibinfo {year} {2021})},\
  \Eprint {http://arxiv.org/abs/2111.09849} {arXiv:2111.09849 [hep-lat]}
  \BibitemShut {NoStop}%
\bibitem [{\citenamefont {d'Enterria}\ \emph {et~al.}(2022)\citenamefont
  {d'Enterria} \emph {et~al.}}]{alphas2022}%
  \BibitemOpen
  \bibfield  {author} {\bibinfo {author} {\bibfnamefont {D.}~\bibnamefont
  {d'Enterria}} \emph {et~al.},\ }\bibfield  {booktitle} {\emph {\bibinfo
  {booktitle} {{2022 Snowmass Summer Study}}},\ }\href@noop {} {\  (\bibinfo
  {year} {2022})},\ \Eprint {http://arxiv.org/abs/2203.08271} {arXiv:2203.08271
  [hep-ph]} \BibitemShut {NoStop}%
\bibitem [{\citenamefont {Boito}\ \emph
  {et~al.}(2021{\natexlab{b}})\citenamefont {Boito}, \citenamefont {Golterman},
  \citenamefont {Maltman}, \citenamefont {Peris}, \citenamefont {Rodrigues},\
  and\ \citenamefont {Schaaf}}]{Boito:2020xli}%
  \BibitemOpen
  \bibfield  {author} {\bibinfo {author} {\bibfnamefont {D.}~\bibnamefont
  {Boito}}, \bibinfo {author} {\bibfnamefont {M.}~\bibnamefont {Golterman}},
  \bibinfo {author} {\bibfnamefont {K.}~\bibnamefont {Maltman}}, \bibinfo
  {author} {\bibfnamefont {S.}~\bibnamefont {Peris}}, \bibinfo {author}
  {\bibfnamefont {M.~V.}\ \bibnamefont {Rodrigues}}, \ and\ \bibinfo {author}
  {\bibfnamefont {W.}~\bibnamefont {Schaaf}},\ }\href {\doibase
  10.1103/PhysRevD.103.034028} {\bibfield  {journal} {\bibinfo  {journal}
  {Phys. Rev. D}\ }\textbf {\bibinfo {volume} {103}},\ \bibinfo {pages}
  {034028} (\bibinfo {year} {2021}{\natexlab{b}})},\ \Eprint
  {http://arxiv.org/abs/2012.10440} {arXiv:2012.10440 [hep-ph]} \BibitemShut
  {NoStop}%
\bibitem [{\citenamefont {Fujikawa}\ \emph {et~al.}(2008)\citenamefont
  {Fujikawa} \emph {et~al.}}]{Belle:2008xpe}%
  \BibitemOpen
  \bibfield  {author} {\bibinfo {author} {\bibfnamefont {M.}~\bibnamefont
  {Fujikawa}} \emph {et~al.} (\bibinfo {collaboration} {Belle}),\ }\href
  {\doibase 10.1103/PhysRevD.78.072006} {\bibfield  {journal} {\bibinfo
  {journal} {Phys. Rev. D}\ }\textbf {\bibinfo {volume} {78}},\ \bibinfo
  {pages} {072006} (\bibinfo {year} {2008})},\ \Eprint
  {http://arxiv.org/abs/0805.3773} {arXiv:0805.3773 [hep-ex]} \BibitemShut
  {NoStop}%
\bibitem [{\citenamefont {Dam}(2021)}]{Dam:2021ibi}%
  \BibitemOpen
  \bibfield  {author} {\bibinfo {author} {\bibfnamefont {M.}~\bibnamefont
  {Dam}},\ }\href {\doibase 10.1140/epjp/s13360-021-01894-y} {\bibfield
  {journal} {\bibinfo  {journal} {Eur. Phys. J. Plus}\ }\textbf {\bibinfo
  {volume} {136}},\ \bibinfo {pages} {963} (\bibinfo {year}
  {2021})}\BibitemShut {NoStop}%
\bibitem [{\citenamefont {d'Enterria}\ and\ \citenamefont
  {Jacobsen}(2020)}]{dEnterria:2020cpv}%
  \BibitemOpen
  \bibfield  {author} {\bibinfo {author} {\bibfnamefont {D.}~\bibnamefont
  {d'Enterria}}\ and\ \bibinfo {author} {\bibfnamefont {V.}~\bibnamefont
  {Jacobsen}},\ }\href@noop {} {\  (\bibinfo {year} {2020})},\ \Eprint
  {http://arxiv.org/abs/2005.04545} {arXiv:2005.04545 [hep-ph]} \BibitemShut
  {NoStop}%
\bibitem [{\citenamefont {Freitas}\ \emph {et~al.}(2019)\citenamefont
  {Freitas}, \citenamefont {Heinemeyer}, \citenamefont {Beneke}, \citenamefont
  {Blondel}, \citenamefont {Dittmaier}, \citenamefont {Gluza}, \citenamefont
  {Hoang}, \citenamefont {Jadach}, \citenamefont {Janot}, \citenamefont
  {Reuter}, \citenamefont {Riemann}, \citenamefont {Schwinn}, \citenamefont
  {Skrzypek},\ and\ \citenamefont {Weinzierl}}]{freitas2019theoretical}%
  \BibitemOpen
  \bibfield  {author} {\bibinfo {author} {\bibfnamefont {A.}~\bibnamefont
  {Freitas}}, \bibinfo {author} {\bibfnamefont {S.}~\bibnamefont {Heinemeyer}},
  \bibinfo {author} {\bibfnamefont {M.}~\bibnamefont {Beneke}}, \bibinfo
  {author} {\bibfnamefont {A.}~\bibnamefont {Blondel}}, \bibinfo {author}
  {\bibfnamefont {S.}~\bibnamefont {Dittmaier}}, \bibinfo {author}
  {\bibfnamefont {J.}~\bibnamefont {Gluza}}, \bibinfo {author} {\bibfnamefont
  {A.}~\bibnamefont {Hoang}}, \bibinfo {author} {\bibfnamefont
  {S.}~\bibnamefont {Jadach}}, \bibinfo {author} {\bibfnamefont
  {P.}~\bibnamefont {Janot}}, \bibinfo {author} {\bibfnamefont
  {J.}~\bibnamefont {Reuter}}, \bibinfo {author} {\bibfnamefont
  {T.}~\bibnamefont {Riemann}}, \bibinfo {author} {\bibfnamefont
  {C.}~\bibnamefont {Schwinn}}, \bibinfo {author} {\bibfnamefont
  {M.}~\bibnamefont {Skrzypek}}, \ and\ \bibinfo {author} {\bibfnamefont
  {S.}~\bibnamefont {Weinzierl}},\ }\href@noop {} {\enquote {\bibinfo {title}
  {Theoretical uncertainties for electroweak and higgs-boson precision
  measurements at fcc-ee},}\ } (\bibinfo {year} {2019}),\ \Eprint
  {http://arxiv.org/abs/1906.05379} {arXiv:1906.05379 [hep-ph]} \BibitemShut
  {NoStop}%
\bibitem [{\citenamefont {Llorente}\ and\ \citenamefont
  {Nachman}(2018)}]{Llorente:2018wup}%
  \BibitemOpen
  \bibfield  {author} {\bibinfo {author} {\bibfnamefont {J.}~\bibnamefont
  {Llorente}}\ and\ \bibinfo {author} {\bibfnamefont {B.~P.}\ \bibnamefont
  {Nachman}},\ }\href {\doibase 10.1016/j.nuclphysb.2018.09.008} {\bibfield
  {journal} {\bibinfo  {journal} {Nucl. Phys. B}\ }\textbf {\bibinfo {volume}
  {936}},\ \bibinfo {pages} {106} (\bibinfo {year} {2018})},\ \Eprint
  {http://arxiv.org/abs/1807.00894} {arXiv:1807.00894 [hep-ph]} \BibitemShut
  {NoStop}%
\bibitem [{\citenamefont {Jezabek}\ and\ \citenamefont
  {Kuhn}(1993)}]{Jezabek:1992sq}%
  \BibitemOpen
  \bibfield  {author} {\bibinfo {author} {\bibfnamefont {M.}~\bibnamefont
  {Jezabek}}\ and\ \bibinfo {author} {\bibfnamefont {J.~H.}\ \bibnamefont
  {Kuhn}},\ }\href {\doibase 10.1016/0370-2693(93)90731-V} {\bibfield
  {journal} {\bibinfo  {journal} {Phys. Lett. B}\ }\textbf {\bibinfo {volume}
  {301}},\ \bibinfo {pages} {121} (\bibinfo {year} {1993})},\ \Eprint
  {http://arxiv.org/abs/hep-ph/9211322} {arXiv:hep-ph/9211322} \BibitemShut
  {NoStop}%
\end{thebibliography}%

\end{document}